\documentclass[12pt, a4paper]{article}
\usepackage{cite}
\usepackage{caption} 
\usepackage[T1]{fontenc}
\usepackage[latin9]{inputenc}
\usepackage[active]{srcltx}
\usepackage{graphicx}
\usepackage{amsmath,amsthm,amssymb,graphicx,mathtools,tikz,hyperref,tabu,enumerate,cancel,braket,float}
\usepackage[export]{adjustbox}

\makeatletter

\newcommand{\tl}{{\tilde{\ell}}}
\newcommand{\te}{{\tilde{e}}}
\newcommand{\tn}{{\tilde{\nu}}}   
\newcommand{\la}{\lambda}
 
\newcommand{\vev}[1]{\langle {#1} \rangle}  
\newcommand{\order}[1]{\mathcal{O}\left({#1}\right)} 
\newcommand{\abs}[1]{\left|{#1}\right|} 
\newcommand{\Lcal}{\mathcal{L}} 
\newcommand{\met}{E_T^{\mathrm{miss}}}

\newcommand{\fbi}{\mathrm{fb}^{-1}}

\@ifundefined{textcolor}{}
{%
 \definecolor{BLACK}{gray}{0}
 \definecolor{WHITE}{gray}{1}
 \definecolor{RED}{rgb}{1,0,0}
 \definecolor{GREEN}{rgb}{0,1,0}
 \definecolor{BLUE}{rgb}{0,0,1}
 \definecolor{CYAN}{cmyk}{1,0,0,0}
 \definecolor{MAGENTA}{cmyk}{0,1,0,0}
 \definecolor{YELLOW}{cmyk}{0,0,1,0}
}

\makeatother

\usepackage{babel}
\begin{document}
\begin{titlepage}

\vskip 1.35cm 
\begin{center}

{\large{\bf 
New Bounds on Light Sneutrino Masses: Rare SUSY Signals
}
}

\vskip 1.5cm 

Linda M. Carpenter$^{a}$\footnote{ 
carpenter.690@osu.edu
}, 
Humberto Gilmer$^{a}$\footnote{
gilmer.30@osu.edu
}, 
Junichiro Kawamura$^{a,b}$\footnote{
kawamura.14@osu.edu
}

\vskip 0.8cm 

{\it 
$^a$ Department of Physics, The Ohio State University, Columbus, Ohio, 43210, USA \\ 
$^b$ Department of Physics, Keio University, Yokohama, 223-8522, Japan  
}
\vspace*{0.8cm} 

\begin{abstract}
We study bounds on a neutral component of a weak doublet scalar lepton. 
A typical example of such a particle is sneutrinos in supersymmetric models. 
Using constraints from invisible Higgs decays, we place a lower bound of approximately $\tfrac{m_h}{2}$. 
We recast a mono-$W/Z$ search, with a hadronic vector \-boson tag in order to bound parameter space 
in the sneutrino--charged slepton mass plane. 
We find a lower bound on sneutrinos in the range of $55--100$ GeV in the 36 $\text{fb}^{-1}$ data set 
depending on the mass of the charged component.   
We propose a sensitivity search in the hadronic mono-$W/Z$ channel for HL-LHC 
and discuss both the discovery potential in case an excess is seen 
and the exclusion limit assuming no excess is seen.   
\end{abstract}
\end{center}
\end{titlepage}

\setcounter{footnote}{0}

\section{Introduction}

Supersymmetry (SUSY) remains the leading candidate for beyond the Standard Model (BSM) physics. 
Lower bounds on colored superpartners are in the TeV range in standard decay scenarios. 
For example, lower bounds on gluinos are in the 2 TeV range~\cite{ATLAS-CONF-2020-002,ATLAS-CONF-2019-040,Sirunyan:2019xwh,Sirunyan:2019ctn}. 
However, lower bounds on electroweak superpartners are not nearly as stringent, 
and many unexplored gaps in parameter space exist. 
Lower bounds on electroweak states are much less stringent. 
For example, bounds on charginos---which are nearly mass degenerate 
with the Lightest Supersymmetric Particle (LSP)---are quite unrestricted~\footnote{
If their masses are so degenerate that charginos decay in detectors, 
they can be probed by searches for disappearing tracks~\cite{Aaboud:2017mpt}.}, in the 100 GeV range, 
and largely come from LEP-II~\cite{Abdallah:2003xe,Tanabashi:2018oca}. 
Among the least constrained sparticles are sneutrinos. 
The current completely generic lower mass bounds on sneutrinos are under $\tfrac{m_Z}2$ and actually come from the
measurement of the $Z$ invisible width at the LEP-I~\cite{Decamp:1991uy,Tanabashi:2018oca}.

In addition, collider signatures for the slepton sector are extremely dependent on the sparticle spectrum. 
Though neutralinos are the canonical candidate for the LSP in many models, 
it is possible for sneutrinos to be the LSP~\cite{Hagelin:1984wv,Ibanez:1983kw}. 
This includes versions of general gauge mediation~\cite{Meade:2008wd,Rajaraman:2009ga,Carpenter:2008he}, Dirac gaugino models~\cite{Fayet:1978qc,Polchinski:1982an,Hall:1990hq,Carpenter:2017xru,Carpenter:2016lgo},  
models with Yukawa textures~\cite{Bernhardt:2008jz}, and models with non-universal Higgs masses \cite{Berezinsky:1995cj,Drees:1996pk,Nath:1997qm,Katz:2009qx}
and others. 
In minimal supersymmetric Standard Model (MSSM) realizations of SUSY models with sneutrino LSP or Next-to-LSP (NLSP), 
the charged left-handed slepton mass lies closely above the sneutrino mass. 
This means that charged sleptons decay to sneutrinos with very soft decay products. 
This makes charged slepton detection in the sneutrino LSP scenario difficult, 
as searches in standard channels~\cite{Aad:2019vnb,Sirunyan:2018nwe}  do not apply. 
In these scenarios both sneutrinos and charged sleptons
appear in the decay chain as missing energy. New search strategies
must be developed to explore this type of spectrum.

In this work we propose new sources for generic lower mass bounds on light sneutrinos. 
We first place a model-independent lower mass bound by investigating couplings in the Higgs sector. 
Through four scalar interactions, the Higgs boson may couple to sneutrinos; 
we then bound the light sneutrino parameter space with the Higgs width constraints.
We then propose a new search strategy for the slepton sector for models with a sneutrino LSP/NLSP 
by employing mono-boson searches. 
Mono-particle searches have been very useful in looking for various BSM phenomena 
including compressed SUSY particles \cite{Anandakrishnan:2014exa,Baer:2014cua,Giudice:2010wb}, and dark matter (DM) searches \cite{Goodman:2010ku,Abercrombie:2015wmb}. 
In this work, we choose to recast LHC dark matter searches for heavy
mono-electroweak bosons \cite{Aaboud:2018xdl} to constrain models with light sneutrinos.
In particular, we here propose a recast the 13 TeV ATLAS hadronic mono-$W/Z$ search: 
$pp \rightarrow \tilde{\ell} \tilde{\ell}^*+ W/Z~(\rightarrow jj)$, 
where $\tl$ is a slepton which contains both sneutrinos $\tn$ and charged sleptons $\te$.  
We then perform a sensitivity search in this channel for 3 $\text{ab}^{-1}$
of the full 14 TeV run at the HL-LHC.

This paper is organized as follows. 
In section~\ref{sec-model}, we review sneutrinos in the MSSM 
and define a simplified model with sneutrino to be used for our phenomenological analysis. 
In section~\ref{sec-pheno}, we study constraints on light sneutrinos. 
At first, we explore the parameter space bound on light sneutrinos from Higgs decays. 
We then discuss the mono-$W/Z$ hadronic search at the LHC.  
Current constraints and future sensitivities on sneutrino-charged slepton mass plane are shown in Section~\ref{sec-rslt}. 
Section~\ref{sec-concl} concludes.

\section{Sneutrino models} 
\label{sec-model}

In this paper, we study the phenomenology of a light $SU(2)_L$ doublet scalar lepton (slepton) 
which may have a non-zero lepton number and is distinguished from Higgs doublets. 
A typical example is doublet (left-handed) sleptons in SUSY extensions of the SM. 
Extra sleptons are also introduced in models for explanations 
for the DM and/or the discrepancy of $(g-2)_\mu$ from the SM prediction~\cite{Bennett:2006fi,Tanabashi:2018oca}.

\subsection{Sneutrino (N)LSP models in MSSM} 
\label{sec-sneuMSSM}

There are many implementations of SUSY models where sneutrinos are the lightest MSSM sparticle. 
Some of the models of sneutrino LSPs were detailed in Ref.~\cite{Katz:2009qx,Arina:2007tm}. 
The parameter space of general gauge mediation also contains many regions 
in which the sneutrino is the NLSP and may decay to a light gravitino~\cite{Rajaraman:2009ga, Carpenter:2008he}.  
Although we will not consider $R$-parity violation here, 
it has been noted that in models with $R$-parity violating operators, 
the renormalization group (RG) 
effects make the sneutrino masses light~\cite{Bernhardt:2008kg, Bernhardt:2008jz}.
The light sleptons might be helpful to explain the long standing discrepancy 
in the anomalous magnetic moment of the muon, $(g-2)_\mu$~\cite{Lopez:1993vi,Chattopadhyay:1995ae,Moroi:1995yh,Endo:2017zrj}.

In this paper, we consider the case that the sneutrino is the lightest particle in the MSSM, 
but is not a sizable fraction of the DM.   
It is well known that the sneutrino LSP is excluded as the DM candidate 
due to the strong constraints from the direct detection~\cite{Goodman:1984dc,Freese:1985qw,Falk:1994es}. 
The $Z$ boson exchange induces the sizable DM-nucleus scattering cross section.    
This constraint can be evaded 
if the relic density of the sneutrino is so tiny that the detection rate is below the current bounds, 
although other DM candidates are required in this case~\footnote{
Mass splitting between the CP-even and -odd states in the sneutrino 
will reduce the direct detection rate 
and some parameter space would be viable~\cite{Hall:1997ah,Cirelli:2005uq,TuckerSmith:2004jv}. 
}. 
Another possibility is that the sneutrino is the NLSP 
and there is a lighter SUSY particle, such as a gravitino~\cite{Moroi:1993mb,Feng:2003xh, Feng:2003uy,Ellis:2003dn,Feng:2004zu} or an axino~\cite{Goto:1991gq,Chun:1992zk,Covi:2001nw,Covi:2004rb,Choi:2011yf}.

In the MSSM, the relevant terms in the scalar potential for doublet sleptons $\tl_i = (\tn_i,\ \te_i )$ 
and the neutral components in the Higgs doublets $h^0_d$, $h^0_u$ are given by 
\begin{align}
 V_{\tilde{\ell}}^\mathrm{MSSM} 
=&\ \sum_{i=1,2,3} \Biggl[m_{L_i}^2 \left( \abs{\tn_i}^2 + \abs{\te_i}^2  \right) \Biggr. \\ \notag 
&\ \Biggl.  +\frac{g^2}{4} \left( \abs{\tn_i}^2 - \abs{\te_i}^2\right) \left(\abs{h^0_d}^2 - \abs{h^0_u}^2 \right) 
  +\frac{g^{\prime 2}}{4} \left( \abs{\tn_i}^2 + \abs{\te_i}^2 \right) \left( \abs{h^0_d}^2 - \abs{h^0_u}^2 \right) 
\Biggr],   
\end{align}
where $g^\prime$, $g$ are respectively the gauge coupling constants of $U(1)_Y$, $SU(2)_L$ 
and $m_{L_i}^2$ is the soft mass squared of the $i$-th doublet slepton.  
The quartic terms in the second line come from the D-term potential. 
Throughout this paper, we shall assume that 
the soft mass $m_{L_i}$ is flavor independent 
and singlet charged sleptons are much heavier than the doublet ones.     
Hereafter, we will omit the flavor index $i$ for simplicity. 
From the latter assumption, 
the left-right mixing in the charged sleptons may be negligible 
due to suppression by the heavy singlet state as well as the small Yukawa couplings.

The left-handed charged sleptons are not split very much from the sneutrinos 
since the charged and neutral states are given by a single soft mass parameter $m_{L}^2$.  
In the absence of left-right mixing in the charged sleptons, 
the mass splitting between the states induced only by the D-term of $SU(2)_L$ and is given by 
\begin{equation}
\label{eq-delm}
\Delta m := m_{\tilde{e}}-m_{\tilde{\nu}} = - \frac{m_W ^2 \cos 2\beta}{m_{\tilde{e}}+m_{\tilde{\nu}}},  
\end{equation}
where $\tan\beta := \vev{h_u^0}/\vev{h_d^0}$ 
and $m_\te$, $m_\tn$ are charged slepton, sneutrino mass, respectively. 
Here, the mass splitting decreases with increasing slepton masses 
and is $\order{10}$ GeV for $\order{100}$ GeV states. 
This relation implies that the mass splitting is at most the $W$ boson mass in the MSSM~\footnote{
The sizable left-right mixing in the charged slepton mass matrix will $decrease$ the mass of the lighter state, 
and it will never increase the mass splitting from the sneutrino 
as long as the sneutrino is lighter than the charged ones. 
Hence, this bound will be hold even if there is sizable left-right mixing. 
}.

\subsection{\textsl{Sneutrinos} in general models}  

{
The search strategies for sleptons discussed in the next section 
will be applicable for a more general class of scalar fields, consisting of a weak doublet with hypercharge $Y = \tfrac{1}{2}$ 
such that a neutral component \textsl{sneutrino} is present. 
This is because we exploit only the invisibility of sneutrinos 
in the SM boson decays and mono-$W/Z$ searches.  
Hence our study will be applicable to, e.g. the inert Higgs doublet~\cite{Deshpande:1977rw}~\footnote{
One big difference is that there is only one inert Higgs doublet 
in the minimal setup, while there are three generations of sneutrinos in the MSSM. Even if sneutrinos decay to a lighter particle  
the limits from our study will be applicable particularly for mass degenerate region.
}. 

For example, 
scalar doublet leptons are introduced to explain the longstanding discrepancy of $(g-2)_\mu$  
and DM~\cite{Agrawal:2014ufa,Baek:2015fea,Kowalska:2017iqv,Calibbi:2018rzv,Kawamura:2020qxo}.  
It is shown that the DM and slepton masses may need to be degenerate, 
so that the direct detection rate is sufficiently suppressed, 
while the relic density is explained by the thermal freeze-out mechanism 
with the help of coannihilation. 
The mono-$Z/W$ search may be a unique way to probe such mass degenerate region 
at the LHC.
In non-SUSY models, quartic couplings to Higgs bosons are not related to the gauge couplings. 
This may allow a mass difference larger than the bound in Eq.\eqref{eq-delm},   
although too large quartic couplings may induce lower Landau pole scales.

Even larger mass splitting may be induced  
by new $SU(2)_L$ triplet states which have trilinear couplings with sleptons, 
\begin{align}
\label{eq-AT}
- \Lcal_T :=  A_T \tl^\dag T \tl   + h.c. , 
\end{align}
where $T$ is a triplet scalar and $A_T$ is a trilinear coupling constant. 
The mass-squared difference in the sleptons is given by $A_T \vev{T^0}$, 
so that the mass difference can be, in principle, much larger than the $W$ boson mass 
by $A_T \gg m_W$.
These types of triplet coupling are induced in the SUSY models with Dirac gauginos~\cite{Fox:2002bu}.   
{More general models may include weak triplets with non-zero VEVs, e.g. in the seesaw mechanisms for light neutrino masses~\cite{Schechter:1980gr,Lazarides:1980nt,Mohapatra:1980yp,Wetterich:1981bx,Mohapatra:1986aw,Mohapatra:1986bd}.
} 
Note that triplets can easily change the electroweak precision observables from the SM values~\cite{Tanabashi:2018oca}.  
This will considerably restrict parameter spaces of models with triplets.

In this work we will take a phenomenological, 
simplified model approach to the light slepton parameter space. 
The slepton potential in the simplified model is defined as 
\begin{align}
\label{eq-Vsv}
 V_{\tl} = m_{\tn}^2 \abs{\tn}^2 + m_{\te}^2 \abs{\te}^2 + A_{\tn} h \abs{\tn}^2 + \frac{1}{2} \lambda_{\tilde{\nu}} h^2 \abs{\tilde{\nu}}^2,   
\end{align}
where $h$ is the physical CP-even scalar of the SM Higgs boson. 
In the MSSM, the effective trilinear and quartic coupling constant $A_\tn$ and $\lambda_\tn$ are given by 
\begin{align}
\label{eq-AMSSM}
 A_{\tn} = \frac{g m_Z^2}{2 m_W} \sin (\alpha + \beta),  
 \quad 
 \lambda_{\tn} = \frac{g^2 m_Z^2}{4 m_W^2} \cos 2\alpha, 
\end{align}
where the neutral complex Higgs bosons are expanded as 
\begin{align}
 \begin{pmatrix}
  h_u^0 \\ h_d^0
 \end{pmatrix}
= 
 \begin{pmatrix}
 \vev{h_u^0} \\ \vev{h_d^0}
 \end{pmatrix}
+\frac{1}{\sqrt{2}} 
\begin{pmatrix}
 \cos\alpha & \sin\alpha \\ -\sin\alpha & \cos\alpha
 \end{pmatrix}
\begin{pmatrix}
 h \\ H
\end{pmatrix}
+\frac{i}{\sqrt{2}} 
\begin{pmatrix}
 \sin\beta_0 & \cos\beta_0 \\ -\cos\beta_0 & \sin\beta_0
 \end{pmatrix}
\begin{pmatrix}
G \\ A 
\end{pmatrix}.  
\end{align}
Here, $H$ ($A$) is a CP-even (-odd) Higgs boson 
and $G$ is a Nambu-Goldstone boson mode which is absorbed by the $Z$ boson. 
{We note that the quartic coupling with the Higgs boson will not play a significant role 
in the mono-$W/Z$ search studied in the next section, 
since the production cross sections involving the Higgs bosons 
are negligible compared to those involving the $W/Z$ bosons. 
We have checked that this is true in the MSSM, 
and hence we will not consider the effect of the quartic coupling in the following analysis. 
}

\section{Phenomenology} 
\label{sec-pheno}

We will discuss constraints on slepton masses in a simplified model defined in Eq.~\eqref{eq-Vsv} 
from invisible Higgs boson decays and LHC searches. 
 In this paper, we assume the following~\footnote{
{As mentioned in Section~\ref{sec-sneuMSSM}, 
 the sneutrino density should comprise a negligible percent of the total DM density 
 if the sneutrino is stable enough to be DM.  }
 }. 
\begin{itemize}
 \item {Sneutrinos and charged sleptons have universal masses and couplings, respectively.}  
 \item {Sneutrinos are stable, or decay to invisible particles.}  
 \item {Charged sleptons exclusively decay to the sneutrinos through the gauge coupling.}   
\end{itemize}
In the MSSM, these will be realized when the doublet sleptons are the lightest MSSM particle 
and the effects of the tau Yukawa coupling are negligible. 
If the tau Yukawa coupling is sizable due to e.g. large $\tan\beta$, 
the tau slepton might be lighter than the others through RG effects. 
Limits obtained under the universal assumption will give conservative limits for the lighter stau scenarios.

\subsection{Higgs invisible decays}
\label{sec-hdecays} 

The current model-independent limit on the sneutrino mass is obtained 
from invisible decay of $Z$ boson~\cite{Tanabashi:2018oca}~\footnote{
The limit of $m_{\tn} > 94$ GeV~\cite{Tanabashi:2018oca,Abdallah:2003xe} is obtained under an assumption of the CMSSM spectrum. 
}. 
We point out that invisible decay of the Higgs boson constrains the sneutrino mass 
in a model-independent way. 
The limit is tightened because of the heavier mass of the Higgs boson than the $Z$ boson.   
Note that no gauge symmetry can forbid the quartic coupling, $\la |\tl|^2 \abs{H}^2$, 
which induces the Higgs decay to a pair of sneutrinos through the effective A-term of $\order{\la v_H}$. 
In fact, the A-term in Eq.~\eqref{eq-AMSSM} is of this order in the MSSM.

The decay width to sneutrinos is given by 
\begin{equation}
    \Gamma(h\rightarrow\tn\tn^*) 
= N_{\tn} \frac{\abs{A_{\tn}}^2}{16\pi m_h}\sqrt{1 - \frac{4 m_{\tn}^2}{m_h^2}},  
\end{equation}
where the $N_{\tn}$ is the number of sneutrinos.  
The branching fraction of the invisible decay to sneutrinos is evaluated as 
\begin{align}
 \mathrm{Br}\left(h\to\mathrm{inv}\right) 
:=&\  \frac{ \Gamma(h\rightarrow\tn\tn^*) }{\Gamma_{h}}  \\ \notag 
\sim&\  390 \times \left(\frac{4.07\ \mathrm{[MeV]}}{\Gamma_h}\right) 
     \left( \frac{ \abs{A_\tn}}{100\ \mathrm{[GeV]} } \right)^2
\times N_{\tn}  \sqrt{1 - \frac{4 m_{\tn}^2}{m_h^2}} ,  
\end{align}
where $\Gamma_h$ is the total decay rate of the SM Higgs boson. 
Hence, the invisible decay tends to dominate over the Higgs decay if it is kinematically allowed~\footnote{
The total decay rate itself is also constrained, $\Gamma_h<$ 0.013 GeV~\cite{Tanabashi:2018oca},  
but this is much weaker than that from the branching fraction of invisible decay, 
since the bound of $0.013$ GeV is much larger than the total width of the SM Higgs boson, 
$4.07\times 10^{-3}$ GeV. 
},  
while the current limit on the invisible decay branching fraction is 0.13~\cite{ATLAS:2020cjb}.
Thus a sneutrino lighter than half of the Higgs mass is excluded 
unless the effective A-term is suppressed as $\order{1\; \mathrm{GeV}}$   
by the small quartic couplings and/or Higgs mixing angle.

\begin{figure}[t] 
\begin{minipage}[c]{0.5\hsize}
\centering
\includegraphics[width=0.95\textwidth]{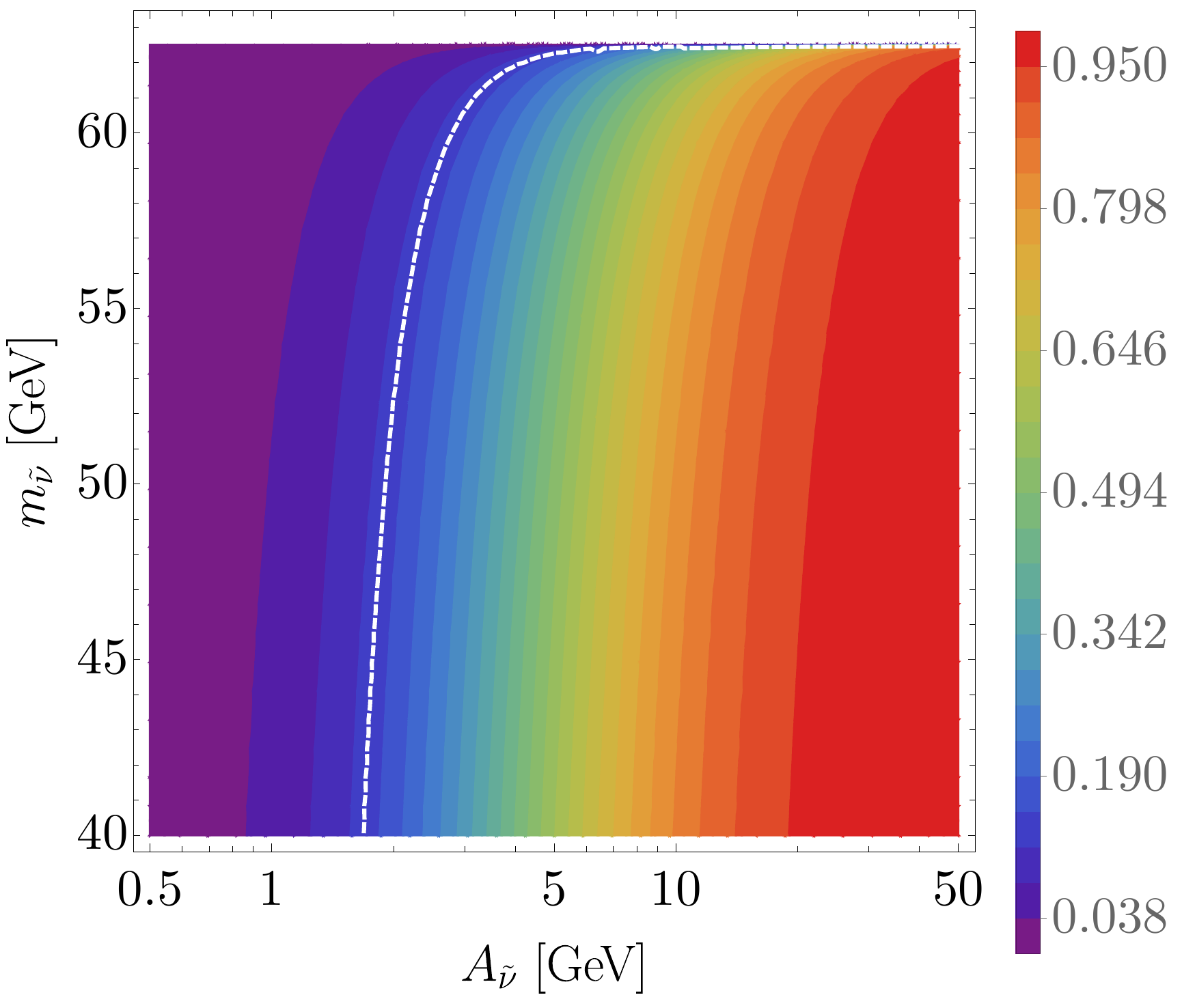} 
\end{minipage}
\begin{minipage}[c]{0.5\hsize}
\centering
\includegraphics[width=0.95\textwidth]{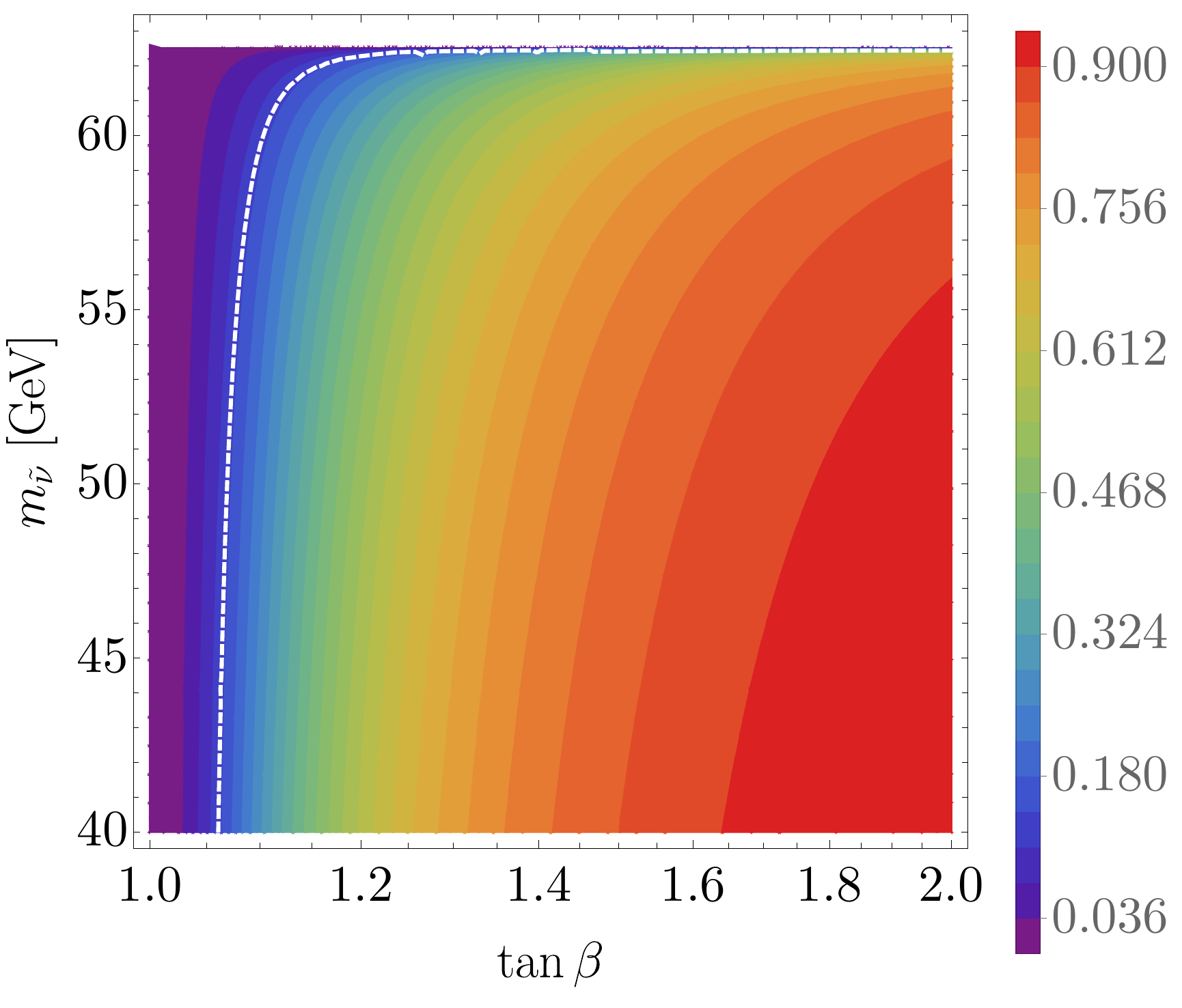} 
\end{minipage}
\caption{\label{fig-hinv}
Contours of Higgs to invisible branching fractions in the general model (left) and in the MSSM (right). 
The white lines show the current limit of $\mathrm{Br}(h\to\mathrm{inv}) = 0.13$~\cite{ATLAS:2020cjb}. 
}
\end{figure}

In the MSSM with the decoupling limit of the heavy Higgs bosons, $\alpha \sim \beta - \pi/2$, 
\begin{align}
\sin(\alpha + \beta) \sim -1 + \frac{2}{1+\tan^2\beta}. 
\end{align}
Thus the invisible decay rate is suppressed by the mixing angle when $\tan\beta \sim 1$, 
otherwise  $m_\tn < \tfrac{m_h}{2}$ is immediately excluded. 
Figure~\ref{fig-hinv} shows exclusion region by the invisible decay of the SM Higgs boson 
in the general model (left) and in the MSSM (right). 
For the MSSM plot, we impose the decoupling limit value, $\alpha = \beta - \pi/2$. 
When $m_\tn \lesssim m_h/2$, the upper bound on the effective A-term is about $2$ GeV 
and this bound corresponds to $\tan\beta \lesssim 1.1$ in the MSSM. 
{Potential problems in the case of such small $\tan\beta \sim 1$ are 
possibly large values for the top Yukawa coupling constant and small tree-level SM-like Higgs mass. 
The top Yukawa coupling constant will blow up below the conventional grand unification scale, $\sim 10^{16}$ GeV, depending on the precise value of the top Yukawa and gauge coupling constants. The existence of the Landau pole would make it difficult to interpret the MSSM as the low-energy theory of conventional Grand Unification Theories. 
The other problem is that the SUSY-breaking masses, particularly top-squark masses, 
need to be heavier than $10^8$ GeV to explain the 125 GeV Higgs boson mass due to the small tree-level contribution~\cite{Bagnaschi:2014rsa}. Hence it would be difficult to justify the lightness of sneutrinos regardless of the heavy squarks. 
If we consider these problems serious, 
it may be necessary to consider $\tan\beta \gtrsim 1.2$, 
and hence the sneutrinos lighter than the half of Higgs boson mass are immediately excluded.
}

\subsection{LHC constraints}
\label{sec-LHC}

\subsubsection{Mass spectra and signals of charged slepton}   

\begin{figure}[t]
\centering
\includegraphics[width=0.75\textwidth]{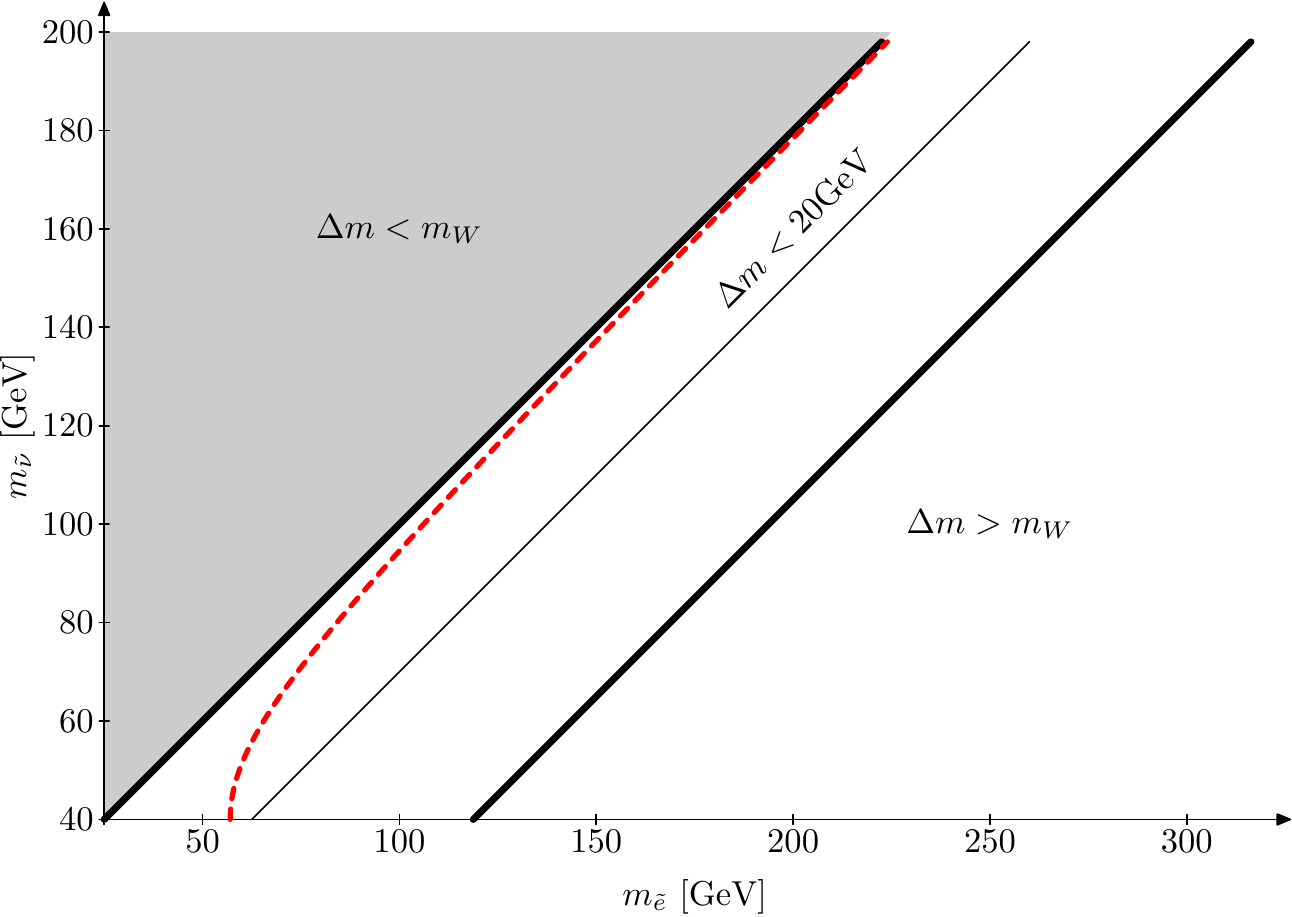}
\caption{Sneutrino-slepton parameter space, showing the various regions explored. 
The region of mass splitting above the red line is consistent with the MSSM 
for reasonable values of $\tan\beta$. 
}
\label{fig:mass_plane}
\end{figure}

In the light sneutrino model,  the charged slepton will decay to a sneutrino through a $W$ boson,  
where this $W$ boson will either be on-shell or off-shell depending on the mass splitting.  
We can then divide the mass parameter space into 3 regions by the mass difference, 
$\Delta m := m_\te - m_\tn$, 
\begin{align}
 (1)~\Delta m \lesssim 20\ \mathrm{GeV},\quad 
 (2)~20\ \mathrm{GeV} \lesssim \Delta m <  m_W,\quad 
 (3)~m_W < \Delta m. 
\end{align}
We visualize these regions in Figure~\ref{fig:mass_plane}. 
In the first region the mass compression is such that any decay products of the off-shell $W$ boson 
resultant from the charged slepton decay will be very soft, 
thus charged sleptons produced in electroweak processes will likely appear in searches as missing energy. 
In the second region, the jets or leptons resultant from the $W$ decay will be hard enough to observe 
but would not be enough to reconstruct as $W$ boson-like object.  
In the third region, the slepton will decay to the sneutrino via an on-shell $W$, 
which will be reconstructed as a $W$ boson-like object.

\begin{figure}[t]
\centering
\includegraphics[width=0.8\textwidth]{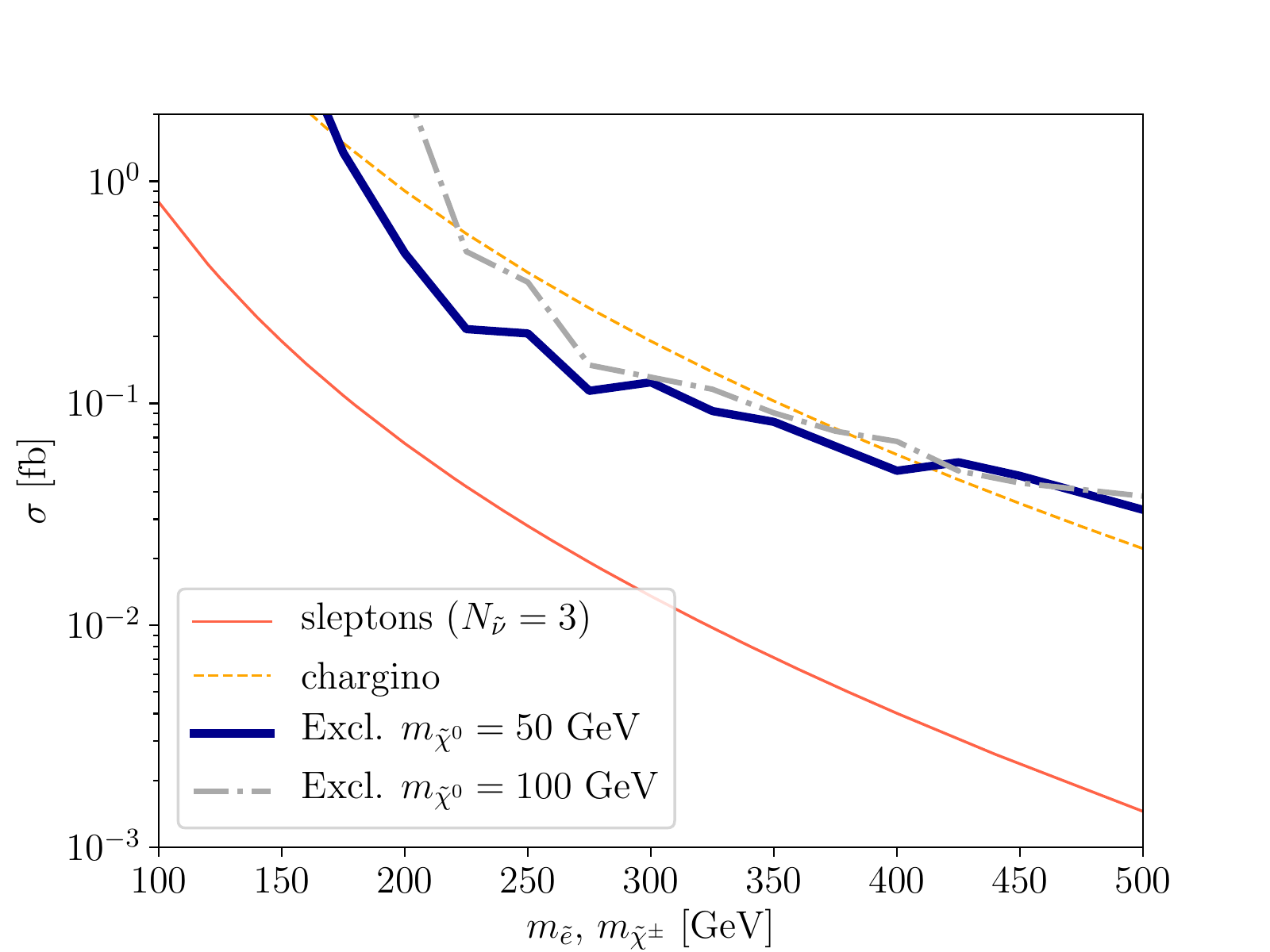}
\caption{Comparison of signal cross section of charged slepton pair 
with experimental limits from the chargino search~\cite{Aad:2019vnb}. 
}
\label{fig-seseprod}
\end{figure}
Although we consider the parameter space where the charged sleptons are heavier than the sneutrinos, 
there might be constraints from the LHC searches for the pair production of charged sleptons, 
\begin{align}
  p p \to \te \te \to W^{(*)} W^{(*)} \tn \tn, 
\end{align}
where the $W$ bosons are off-shell in the first and second regions and are on-shell in the third region~\footnote{ 
Hereafter, we will omit symbols for the conjugation of sleptons $^*$, 
but we use this symbol for off-shell particles.}.
This signal is the same as that from a pair production of charginos~\cite{Aad:2019vnb}.  

Figure~\ref{fig-seseprod} shows the signal cross sections of doublet charged sleptons (red solid)
and wino-like charginos (yellow dashed) at $\sqrt{s} = 13$ TeV. 
The production cross sections were calculated 
in Refs.~\cite{Bozzi:2007qr,Fuks:2013vua,Fuks:2013lya,Fiaschi:2018hgm,Beenakker:1999xh} 
and in Refs.~\cite{Fuks:2012qx,Fuks:2013vua}, respectively.  
Here, the slepton production cross section is calculated for the three-flavor sneutrinos. 
The blue thick and gray dashed lines are the exclusion limits on the signal cross sections 
where the neutral particle is 50 and 100 GeV, respectively. 
The values are shown in the supplemental material of Ref.~\cite{Aad:2019vnb}. 
The efficiency times acceptance factor ($\epsilon\times \mathcal{A}$) for the signal cross section is set to $10^{-3}$ such that a chargino within about $[200, 400]$ GeV is excluded by the exclusion limit for $m_{\tilde{\chi}^0} = 100$ GeV.   
Assuming the same efficiency times acceptance factor, 
we see that the charged slepton pair production will not be detectable at the LHC due to the too small production cross section.

\subsubsection{Sneutrino signals}   

\begin{figure}[t]
\centering
\includegraphics[width=0.8\textwidth]{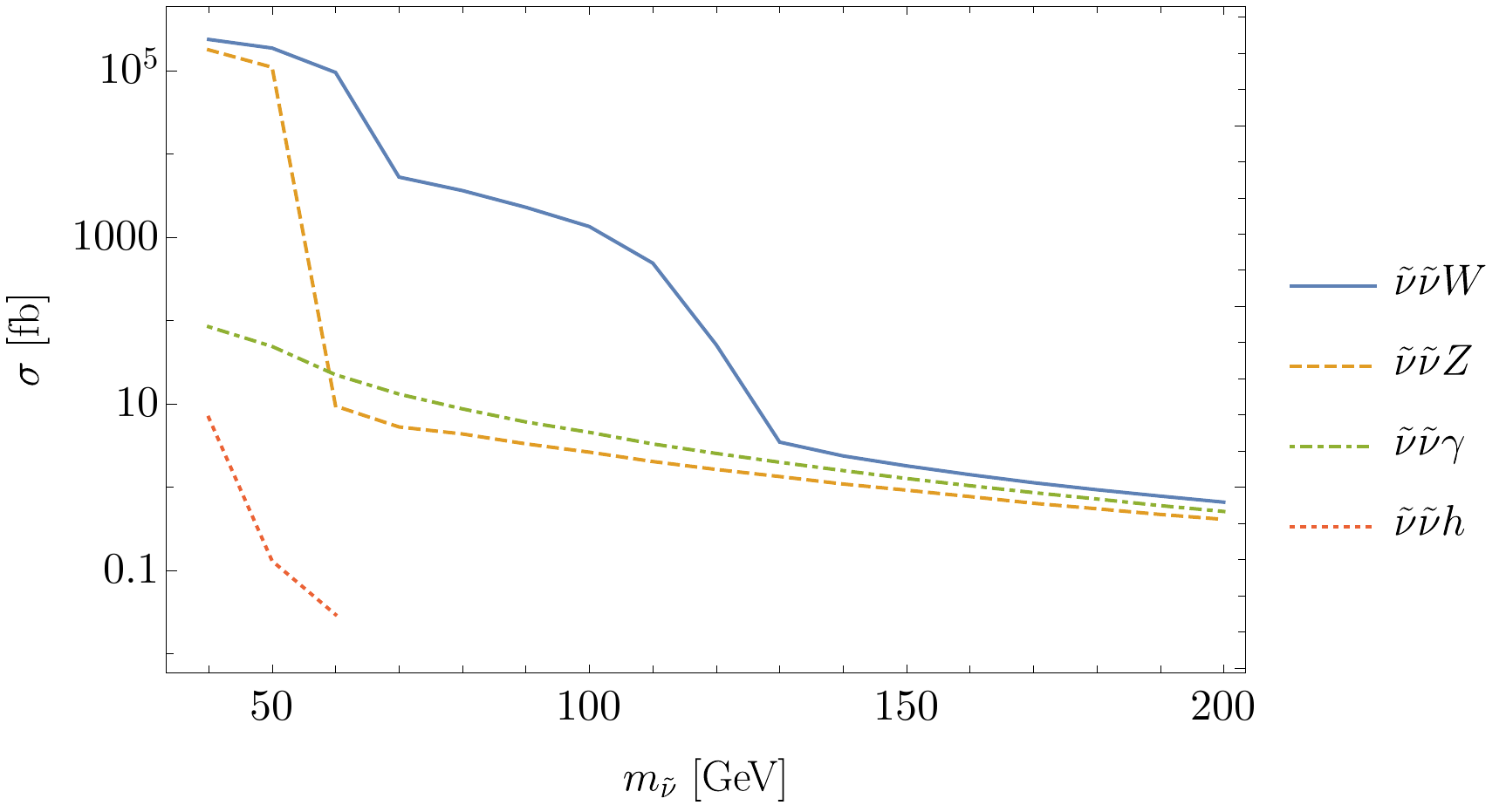}
\caption{Production cross sections for sneutrino pair in association with the electroweak bosons. The charged slepton mass is fixed at $m_\te = 200$ GeV.}
\label{fig:cross_sections}
\end{figure}
As light sneutrinos appear as missing energy in the detector, 
events with final-state sneutrinos should be produced in association with a visible particle.  
As such, we choose to focus on LHC searches 
where sneutrinos are produced in association with a single boson. 
The mono-jet analysis, whose events result from an initial-state-radiation (ISR) gluon, clearly has the largest production cross section; 
however since there is a very large background the search will likely not have a great sensitivity 
into the sneutrino parameter space. 
For example, projections for the 3 $\mathrm{ab}^{-1}$ run of LHC using the monojet search 
can bound electroweak gauginos at about 100 GeV~\cite{Baer:2014cua} 
and slepton production cross sections are much smaller. 
The likely signal-to-background ratio does not portend well for this channel.

We thus turn to the production of sneutrinos and charged sleptons in association with the electroweak bosons. 
In Fig.~\ref{fig:cross_sections}, we plot cross sections versus sneutrino masses for sneutrinos pair produced 
in association with various electroweak bosons, the $W$, $Z$, photon and Higgs boson. 
In this plot, the charged sleptons are fixed at 200 GeV. 
Note that in the mono-Higgs case, 
the Higgs boson is produced as final-state-radiation (FSR) off of an internal gauge boson line, 
or off of a sneutrino line due to the four scalar interaction mentioned previously. 
The photon is produced as initial state radiation. 
$W$ production, however, may originate from ISR, FSR, or from an on-shell particle decay. 
We see that the $W$ boson production dominates the production cross sections.  
Balancing between production cross sections, and anticipated signal-to-background ratios, 
we have chosen to recast the ATLAS mono-boson analysis~\cite{Aaboud:2018xdl} 
which searches for an on-shell hadronically tagged $W$ or $Z$ boson produced 
in association with large missing energy, $pp \rightarrow  \met + W/Z (\to jj)$.

We will now discuss how this search applies to our three regions of parameter space 
in the sneutrino--charged slepton mass plane:
\begin{enumerate}
    \item In the mass-degenerate region both the sneutrinos and the softly decaying charged sleptons will appear as missing energy in the search. This means that not only does sneutrino pair production contribute to the overall production cross section, but so does the production of a sneutrino--charged slepton pair, and the pair production of softly decaying charged sleptons. The relevant processes are $p p \to W/Z + \tl \tl$, 
where $\tl$ contains both $\tn$ and $\te$. 
    \item In the intermediate region, decays of the charged slepton are hard enough for decay products to be detected; however, since the decay products of the resultant $W$ boson are off-shell, the events will not have $W/Z$-tagged jets. In this region, the only contributing process will be the pair production of sneutrinos in association with a $W$ or $Z$ boson: $pp \rightarrow W/Z+\tilde{\nu} \tilde{\nu}$.
    \item In the large mass splitting region, two types of events contribute to the overall cross section. One is of course the pair production of sneutrinos in association with a $W$ boson as above $ p p \rightarrow W/Z+\tilde{\nu} \tilde{\nu}$. However, the main production cross section results from the production of one sneutrino and one charged slepton. The charged slepton then undergoes a decay to a sneutrino through an on-shell $W$, which will be caught in the hadronic vector boson tag. The relevant process is $pp\rightarrow \te \tn \rightarrow W \tn\tn$.
\end{enumerate}

In considering the large mass-splitting region we may further understand the features 
of Fig.~\ref{fig:cross_sections}. 
In the $W$ and $Z$ associated channels 
we see that the production drops after the sneutrino mass passes $\tfrac{m_Z}/{2}$
as there is an enhancement for light sneutrinos produced by the decay of an on-shell $Z$ boson. 
There is an additional feature in the mono-$W$ channel. 
With the charged slepton mass fixed at 200 GeV, we see that for sneutrino masses lighter than about 120 GeV,
the slepton mass splitting is large and the $W \tn \tn$ cross section is dominated 
by the production of a charged slepton-sneutrino pair. 
The slepton decays to the on-shell $W$ and a sneutrino. This production cross section is quite high. 
Once the sneutrino mass becomes large enough, the on-shell $W$ cannot be produced by the slepton decay. 
Instead the on-shell vector boson must be produced as ISR/FSR.  
This production has an implicitly lower cross section.

\begin{figure}[t] 
\begin{minipage}[c]{0.5\hsize}
\centering
\includegraphics[width=0.95\textwidth]{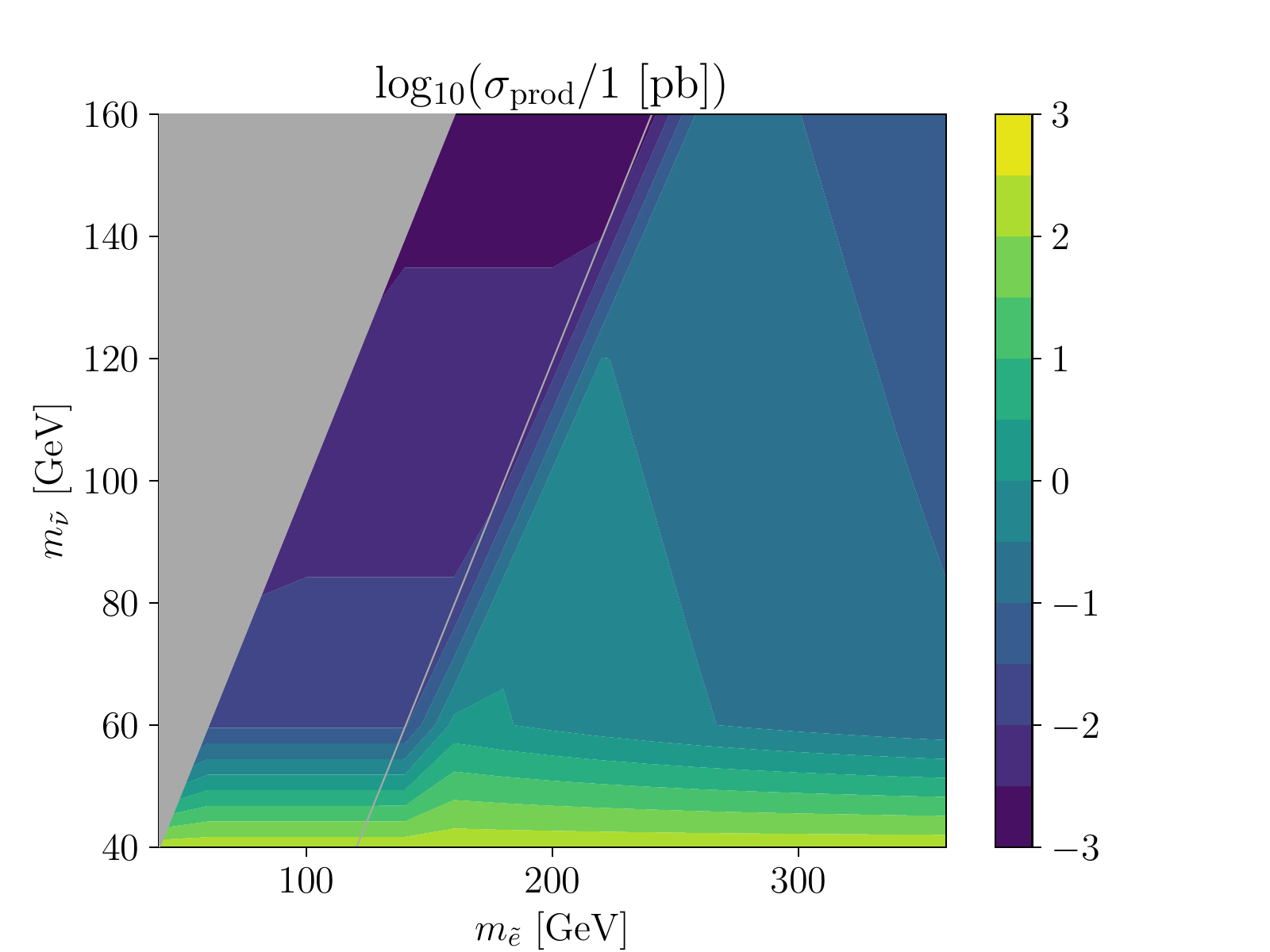} 
\end{minipage}
\begin{minipage}[c]{0.5\hsize}
\centering
\includegraphics[width=0.95\textwidth]{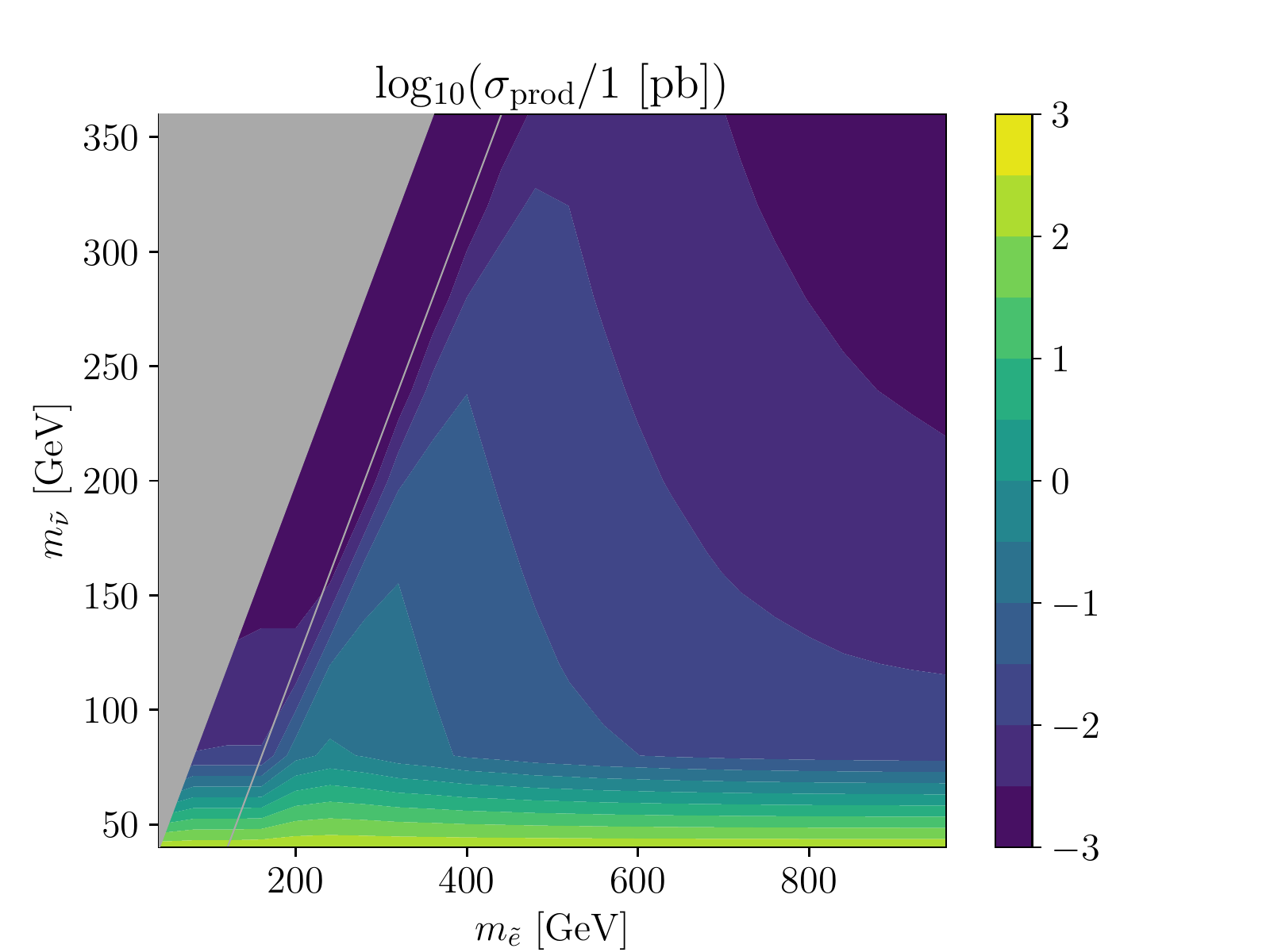} 
\end{minipage}
\caption{\label{fig-xsection-contour}
Contour plot of production cross sections of $pp\to W/Z + \tn \tn$ 
at $\sqrt{s} = 13$ TeV (left) and $14$ TeV (right) in the ($m_\te$, $m_\tn$) plane. 
}
\end{figure}

Figure~\ref{fig-xsection-contour} shows the total production cross sections $\tilde{\nu}\tilde{\nu} + W/Z$ 
in the sneutrino--charged slepton mass plane.
The production cross section is dominated by $\tn\tn$ production in associated 
with initial-/final-state W/Z boson where $m_{\tn} \lesssim 60$ GeV.   
For heavier sneutrinos, the production cross section is smaller than $\order{0.01}$ pb 
in the off-shell $W$ region, 
while it can be $\order{0.1}$ pb in the on-shell $W$ region due to the $\te\tn$ production. 
The cross section slowly decreases as the charged slepton mass increases. 
The charged slepton mass dependence of the cross section is milder than that of the pair production. 
In the next section, 
we describe our recast of the ATLAS hadronic-tagged mono-$W/Z$ search 
and set limits in the sneutrino--charged slepton mass plane.

\subsubsection{Mono $W$/$Z$ Search}

\begin{table}[t]
 \centering
\caption{\label{tab-monoW}
Event selection criteria in the mono $W/Z$ search~\cite{Aaboud:2018xdl}.  
The symbols $j$ and $J$ are the small-$R$ and large-$R$ jets, respectively. 
$j_i$'s are the small-$R$ jets ordered by their $p_T$ in decreasing order. 
Angles are defined in radians. See the text for details. 
}
\small
\begin{tabular}[t]{c|cccc|cc}\hline\hline
                     & \multicolumn{4}{c|}{Merged topology} & \multicolumn{2}{|c}{Resolved topology} \\   \hline\hline 
   $\met$      & \multicolumn{4}{c|}{$>250$ GeV }       & \multicolumn{2}{|c}{$>150$ GeV} \\   
Jets, leptons & \multicolumn{4}{c|}{$\ge 1$J, 0$\ell$}   & \multicolumn{2}{|c}{$\ge 2j$, 0$\ell$} \\   
 b-jets           & \multicolumn{4}{c|}{no b-tagged jets outside of $J$}&\multicolumn{2}{|c}{$\le 2$ b-tagged small-$R$ jets} \\   
\hline 
                     & \multicolumn{6}{c}{$\Delta \phi (\vec{E}_T^\mathrm{miss},\ J\ \mathrm{or}\ jj ) > 2\pi/3$} \\   
  Multijet&\multicolumn{6}{c}{$\min_{i=1,2,3}\left[\Delta\phi(\vec{E}_T^\mathrm{\ miss}\ j_i)\right]>\pi/9$} \\ 
  suppression& \multicolumn{6}{c}{$\abs{\vec{p}_T^\mathrm{\ miss}} > 30$ GeV or $\ge 2$ b-jets} \\ 
           & \multicolumn{6}{c}{$\Delta \phi (\vec{E}_T^\mathrm{\ miss}, \vec{p}_T^\mathrm{\ miss} ) < \pi/2$}\\ 
\hline 
Signal       & \multicolumn{4}{c|}{} & \multicolumn{2}{|c}{$p_T^{j_1} > 45$ GeV} \\   
properties & \multicolumn{4}{c|}{} & \multicolumn{2}{|c}{$\sum_i p_T^{j_i} > 120$ ($150$) GeV for 2 ($\ge 3$) jets} \\
\hline \hline  
Signal region & 0b-HP & 0b-LP & 1b-HP & 1b-LP  & 0b-Res & 1b-Res \\ \hline 
 $J$ or $jj$&HP & LP & HP &LP & \multicolumn{2}{c}{$\Delta R_{jj} < 1.4$ and $m_{jj} \in [65, 105]$ GeV} \\
b-jet              & no b-jet &   no b-jet &   1 b-jet &   1 b-jet &   no b-jet &   1 b-jet \\ 
\hline\hline 
\end{tabular}
\normalsize
\end{table}

\begin{figure}[t] 
\begin{minipage}[c]{0.5\hsize}
\centering
\includegraphics[width=0.95\textwidth]{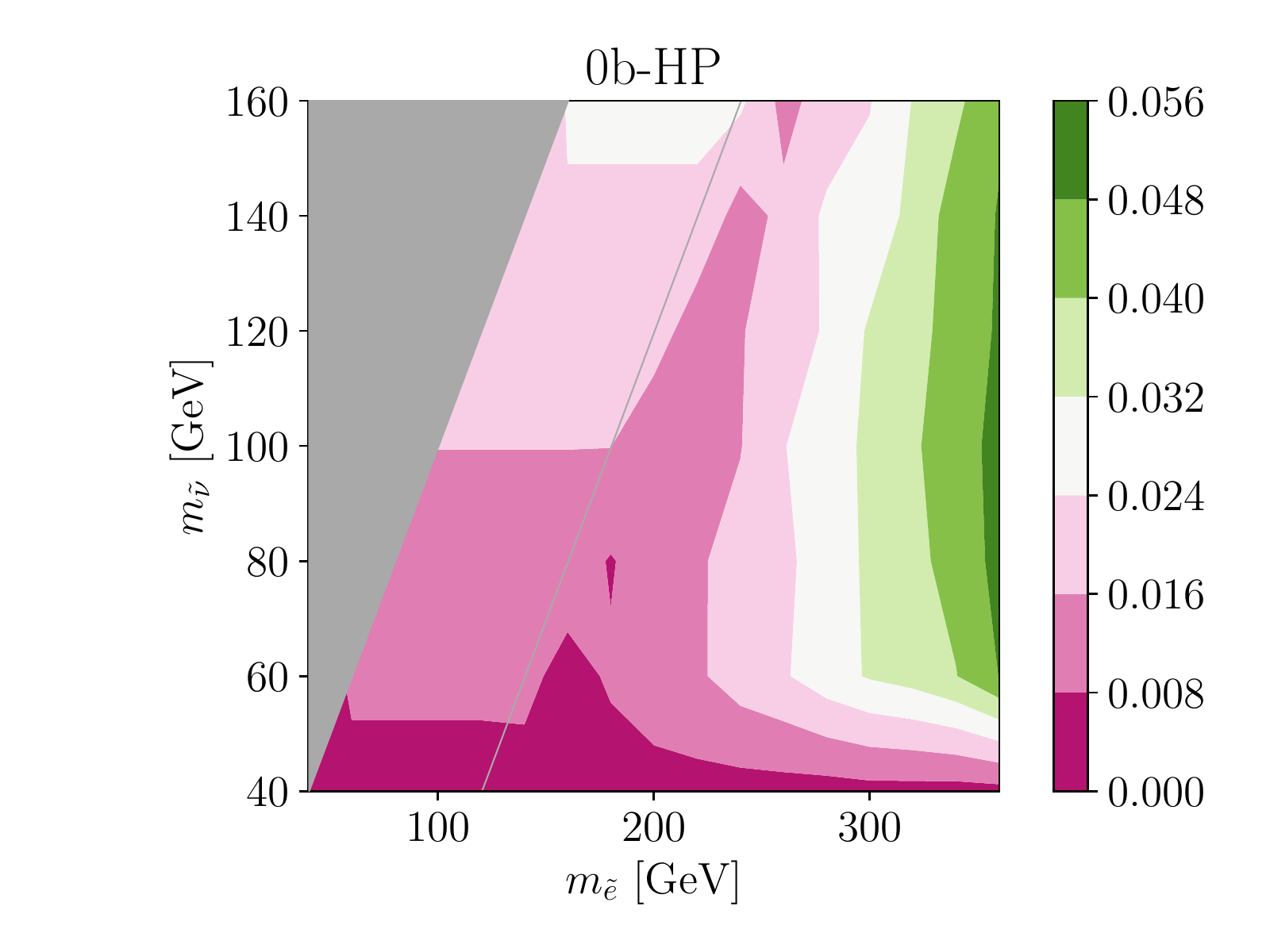} 
\end{minipage}
\begin{minipage}[c]{0.5\hsize}
\centering
\includegraphics[width=0.95\textwidth]{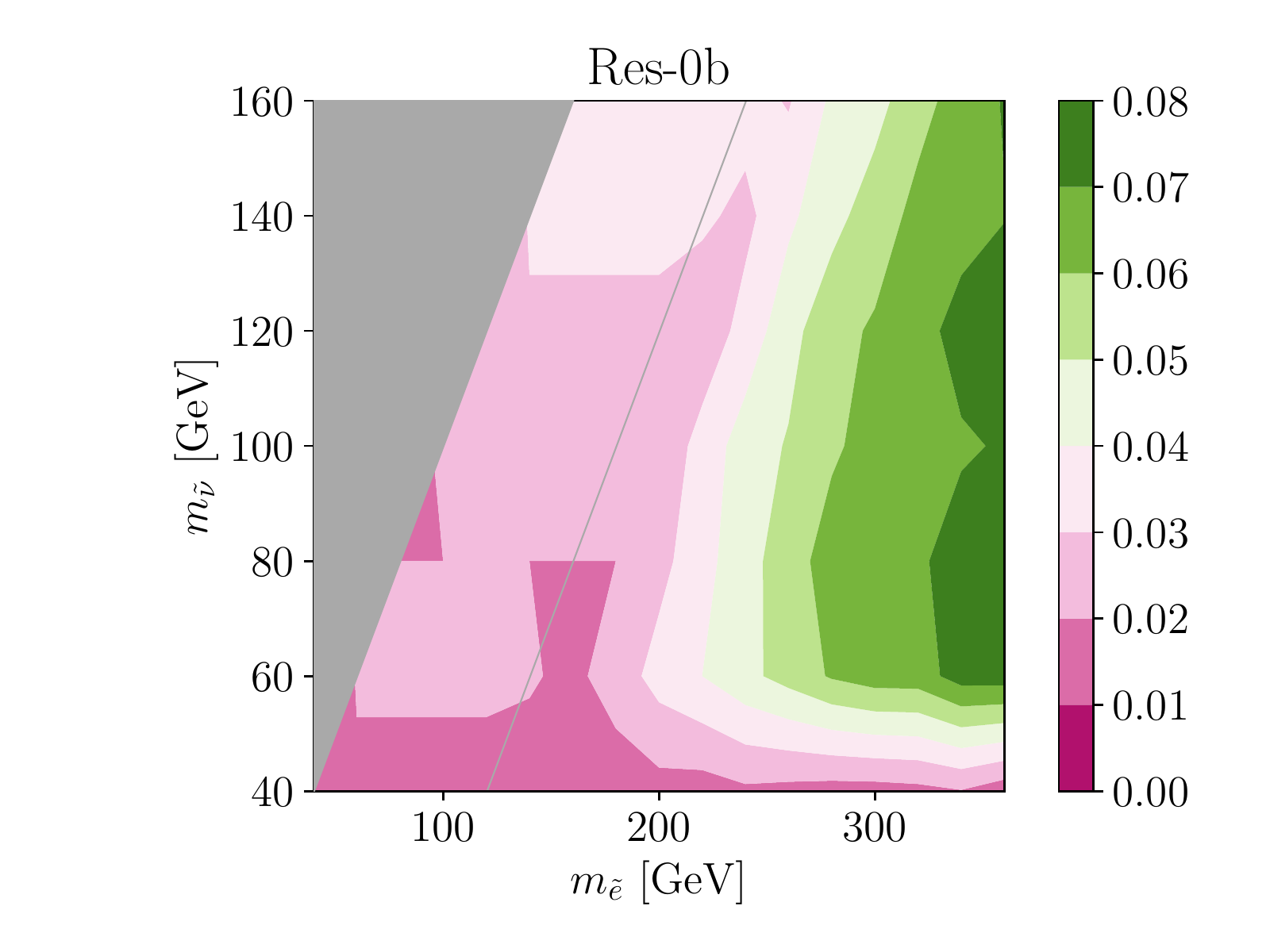} 
\end{minipage}
\begin{minipage}[c]{0.5\hsize}
\centering
\includegraphics[width=0.95\textwidth]{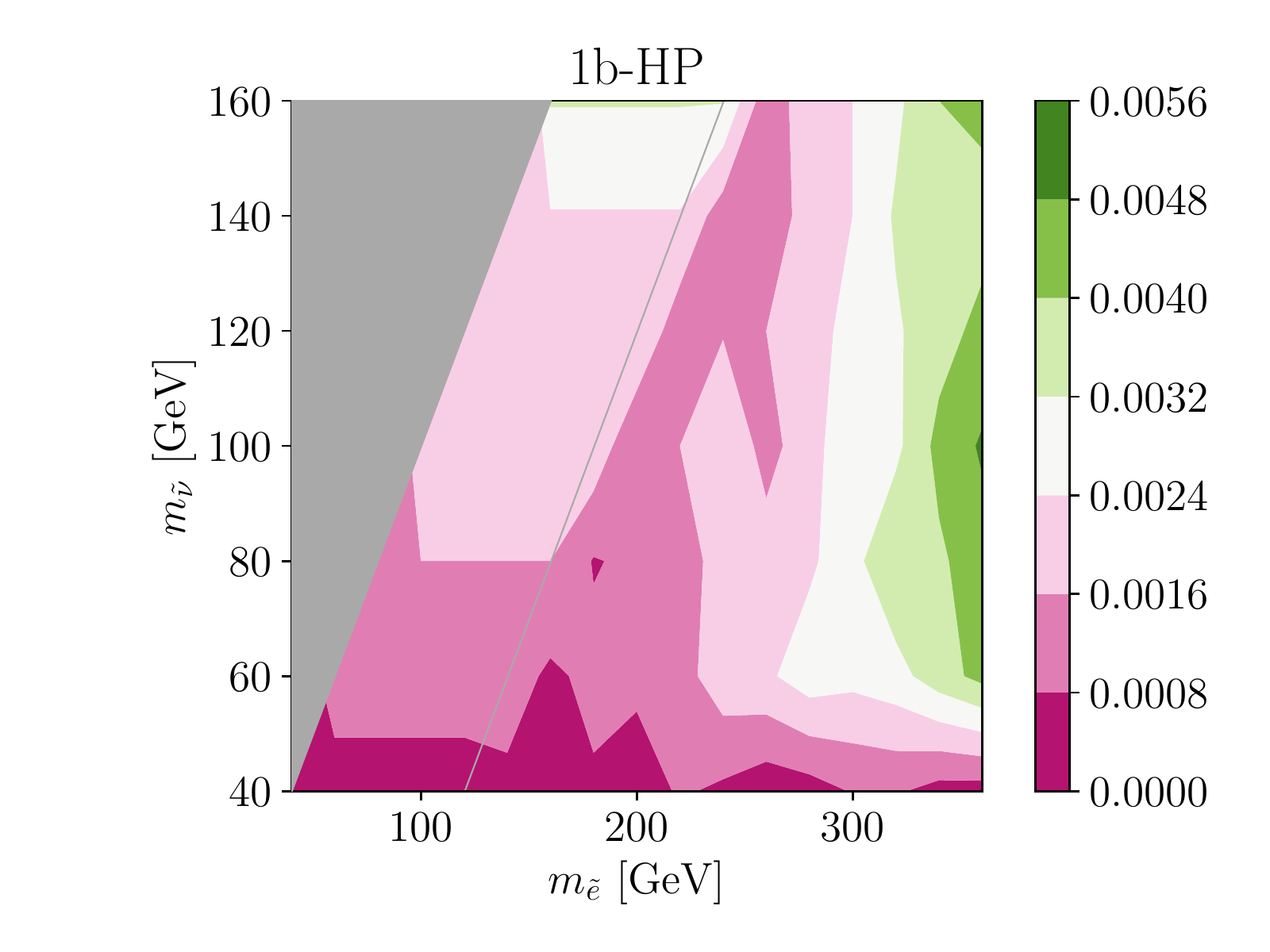} 
\end{minipage}
\begin{minipage}[c]{0.5\hsize}
\centering
\includegraphics[width=0.95\textwidth]{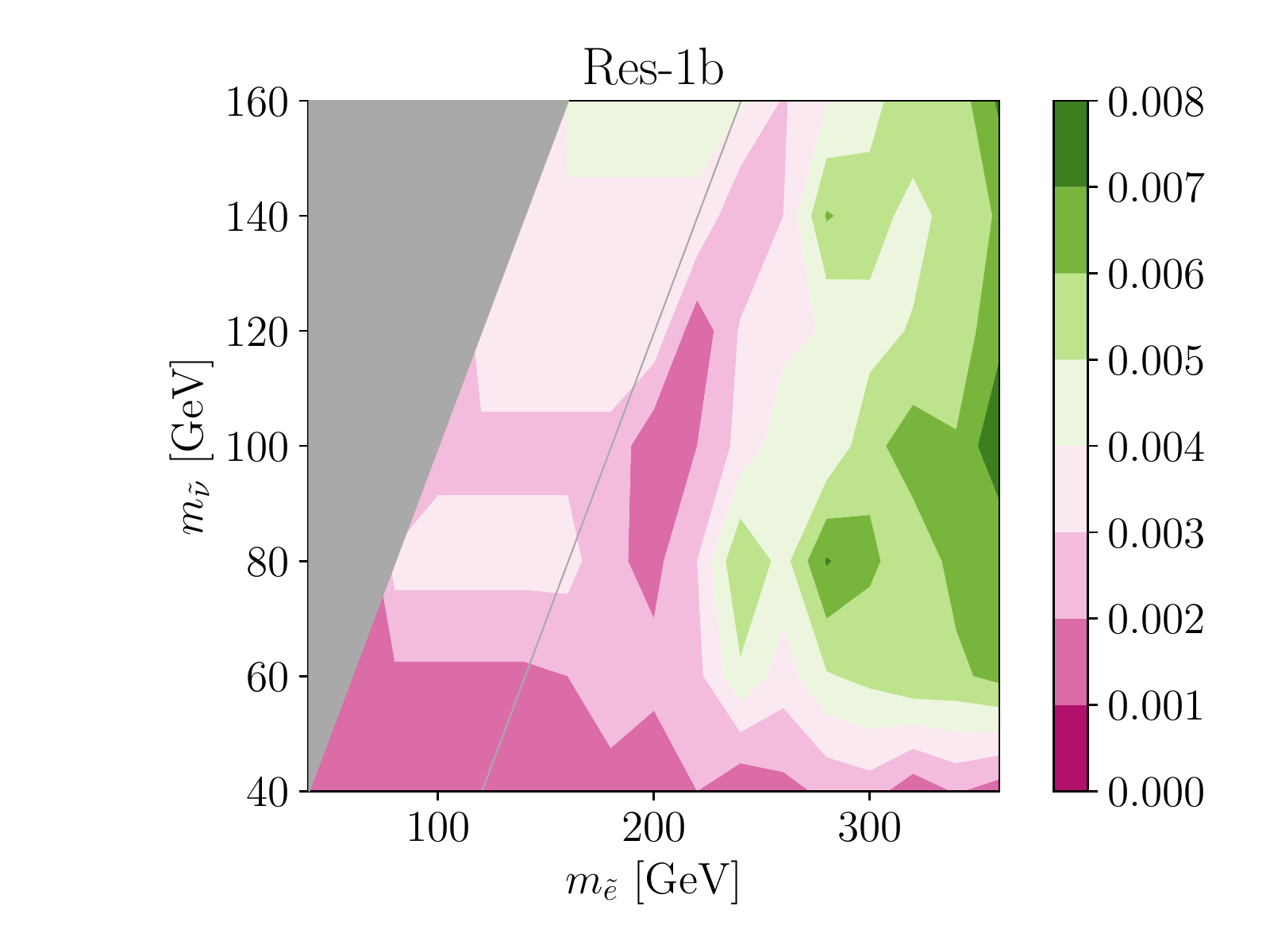} 
\end{minipage}
\caption{\label{fig-efficiency13}
The efficiencies of the sneutrino production in associated with a $W/Z$ boson 
in the 0b-HP (top-left), Res-0b (top-right), 1b-HP (bottom-left) and Res-1b (bottom-right) 
at $\sqrt{s} = 13$ TeV. 
}
\end{figure}

\begin{figure}[th] 
\begin{minipage}[c]{0.5\hsize}
\centering
\includegraphics[width=0.95\textwidth]{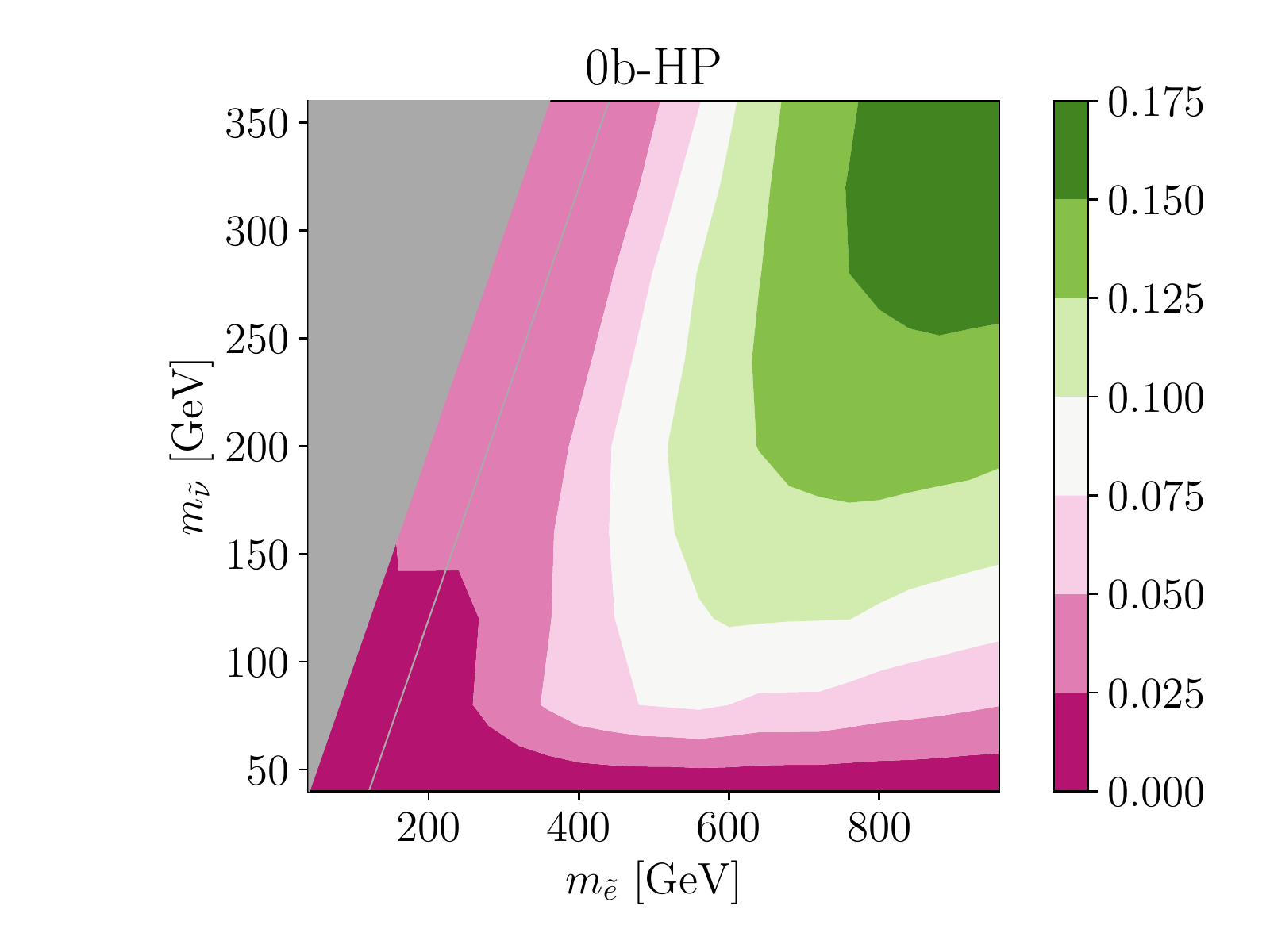} 
\end{minipage}
\begin{minipage}[c]{0.5\hsize}
\centering
\includegraphics[width=0.95\textwidth]{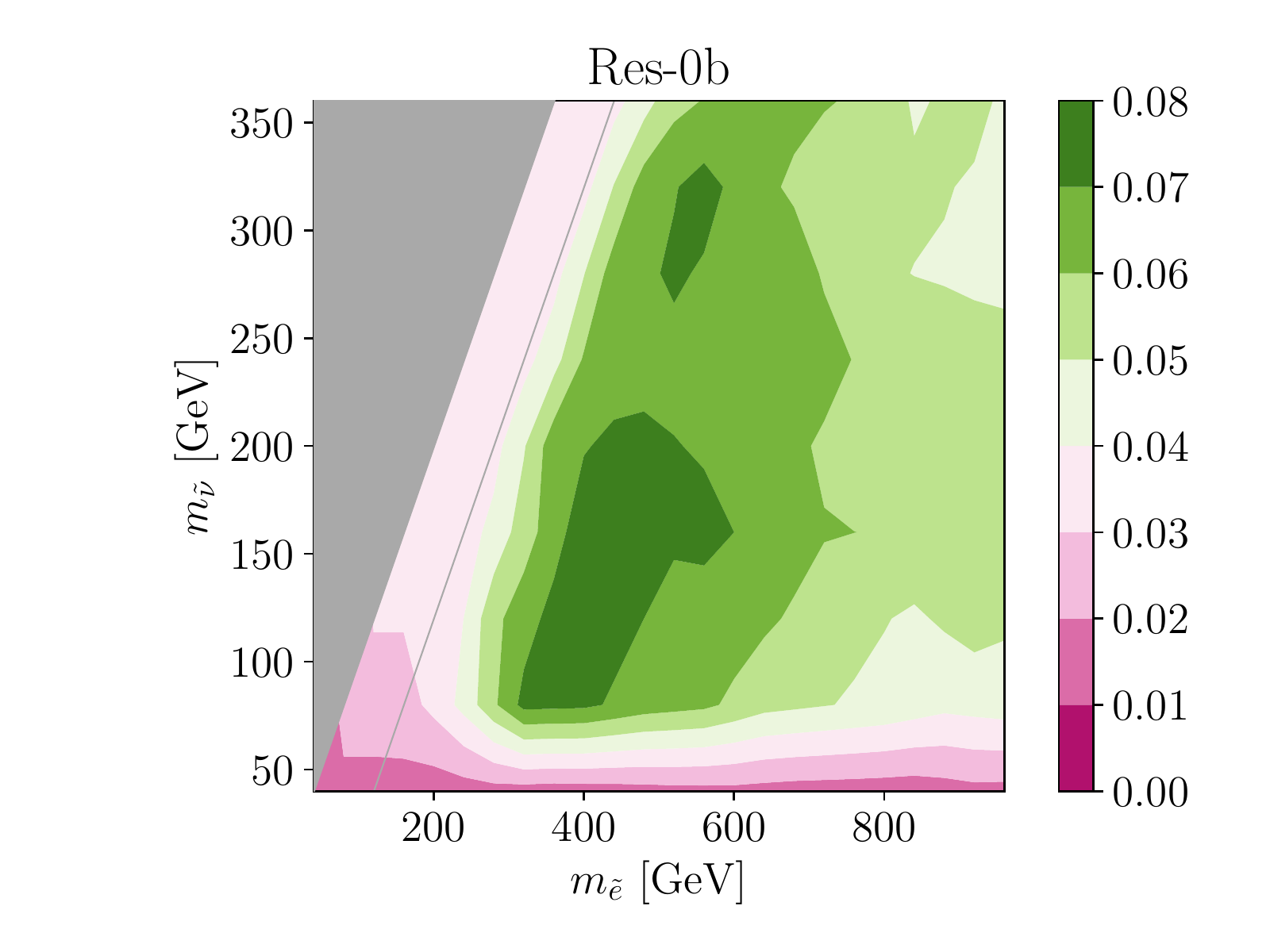} 
\end{minipage}
\begin{minipage}[c]{0.5\hsize}
\centering
\includegraphics[width=0.95\textwidth]{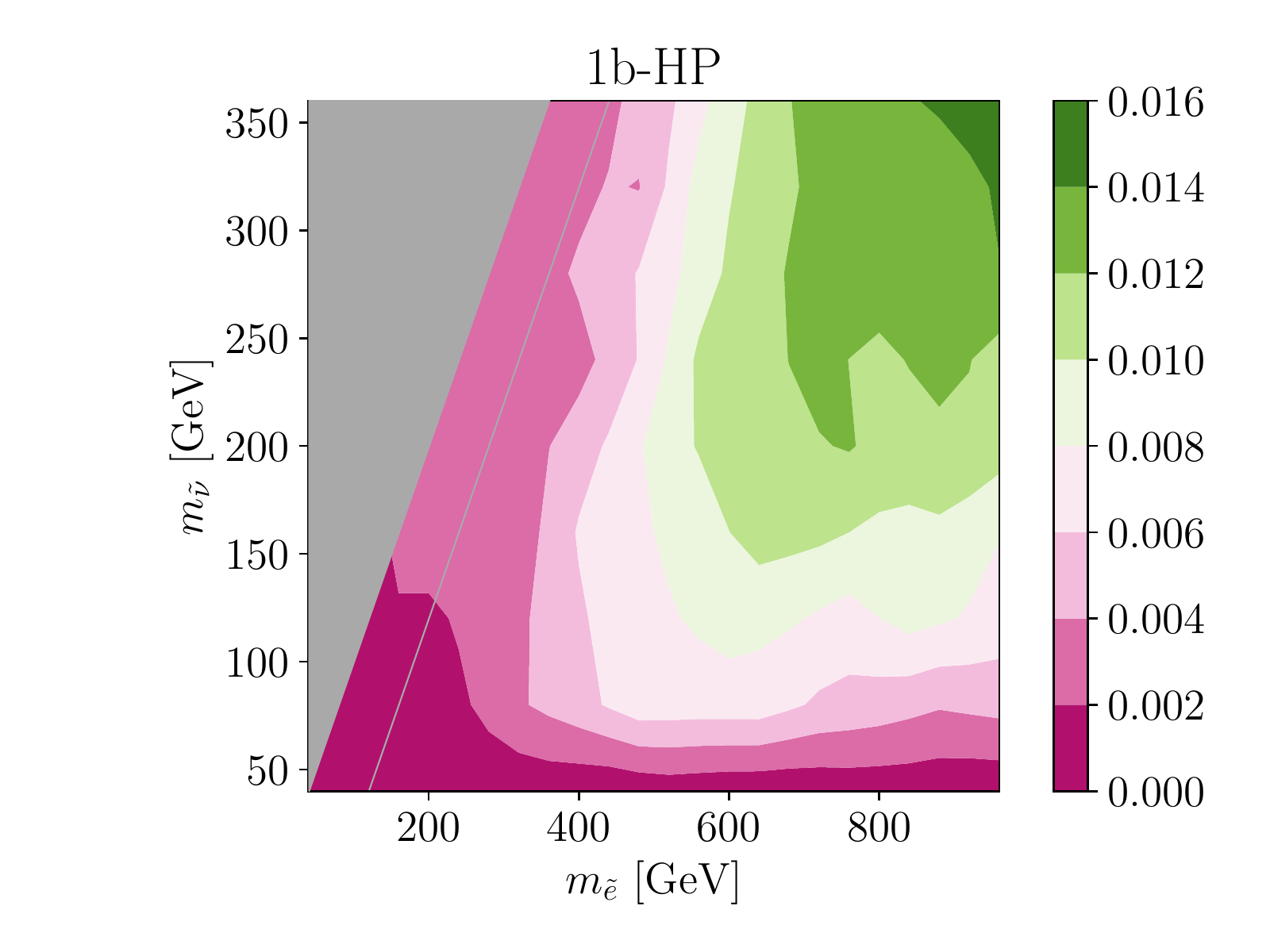} 
\end{minipage}
\begin{minipage}[c]{0.5\hsize}
\centering
\includegraphics[width=0.95\textwidth]{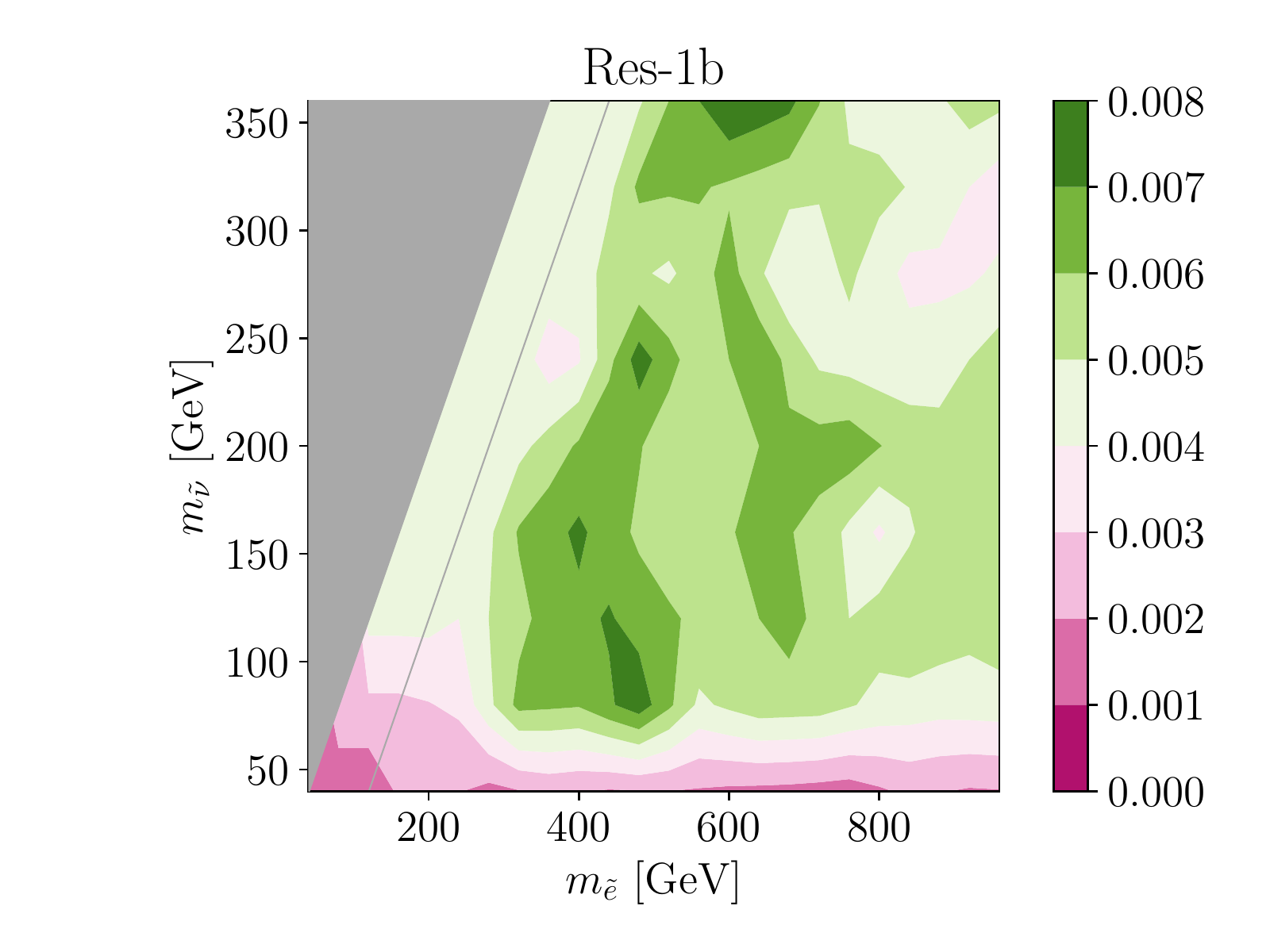} 
\end{minipage}
\caption{\label{fig-efficiency14}
The efficiencies of the sneutrino production in associated with a $W/Z$ boson 
in the 0b-HP (top-left), Res-0b (top-right), 1b-HP (bottom-left) and Res-1b (bottom-right) 
at $\sqrt{s} = 14$ TeV. 
}
\end{figure}

ATLAS has searched for a single on-shell hadronically decaying vector boson 
produced in association with invisible particles~\cite{Aaboud:2018xdl}. 
This search was originally conceived as a search for fermionic models of DM; 
however we may recast this search in order to constrain the light sneutrino scenario.

The event selection criteria in the mono-$W/Z$ search are summarized in Table~\ref{tab-monoW}. 
The search looks for events with large missing energy that contain 
a large-$R$ jet (merged topology) or two jets (resolved topology) 
with dijet invariant mass falling in a window around the $W/Z$ boson mass.  
In both topologies, large missing transverse energy ($\met$) is required 
and any events with reconstructed leptons are rejected. 
In order to suppress multijet backgrounds, 
azimuthal separation between the $\met$ vector and the large-$R$ jet (the two highest-$p_T$ jets system) 
is required to be larger than $2\pi/3$ in the merged (resolved) topology. 
The azimuthal separations between the $\met$ vector and the leading 3 jets are also required. 
In addition, track-based missing transverse momentum, 
$\vec{p}^{\mathrm{miss}}_T$ defined as the negative vector sum of the transverse momenta of tracks with 
$p_T > 0.5$ GeV and $\abs{\eta} < 2.5$, is required to be larger than 30 GeV and 
its azimuthal angle is close to that of the calorimeter-based $\met$ within $\pi/2$.

In the merged topology, any $b$-tagged jets outside of the large-$R$ jet is rejected 
on top of the above requirements.  
The signal regions in this topology are separated by the number of $b$-tagged jets 
and the purity of the large-$R$ jet to be tagged as origination from a hadronic vector boson decay.  
Since the selection criteria in the analysis~\cite{Aaboud:2018xdl} is adjusted 
such that the efficiency is of 50 $\%$ independent of jet $p_T$,    
we simply assume that a half of events with the large-$R$ jet is classified into the high-purity (HP) regions, 
and the rest of events is classified into the low-purity (LP) regions. 
Since the sneutrino pair production is mostly associated with $W$ boson, 
there are very few events with 2 $b$-tagged jets. 
In fact, we found that the number of events in the signal region with 2 $b$-tagged jets 
is always less than the experimental limit.  
Hence, we will study the signal regions with up to one $b$-tagged jet. 
Thus, we will study the four signal regions, 0b-HP, 0b-LP, 1b-HP and 1b-LP.

In the resolved topology, 
the highest-$p_T$ jet is required to be $p_T > 45$ GeV 
and the sum of $p_T$ of the two (three) leading jets is required to be larger than 120 (150) GeV. 
In addition, 
the angular separation $\Delta R := \sqrt{(\Delta \phi)^2 + (\Delta \eta)^2}$ between the two leading jets 
and the invariant mass of the two leading jets   
are required to be smaller than 1.4 and within a range $[65, 105]$ GeV, respectively.   
Events in this topology is classified into three signal regions by the number of $b$-tagged jets. 
We will study the two signal regions, 0b-Res and 1b-Res.

We have generated signal events using {\sc MadGraph5$\,2.7.2$}~\cite{Alwall:2014hca}, 
showered events with {\sc Pythia 8.2.4.5}~\cite{Sjostrand:2007gs}, 
and ran events through the fast detector simulator {\sc Delphes3}~\cite{deFavereau:2013fsa}.  
We used the default ATLAS card for the detector simulation, 
but we add the large-$R$ jet with $R = 1.0$ on top of the small-$R$ jet with $R = 0.4$ 
reconstructed by using anti-$k_T$ jet clustering algorithm~\cite{Cacciari:2008gp,Cacciari:2011ma}. 
The trimming algorithm~\cite{Krohn:2009th} is applied to the large-$R$ jets 
as in the analysis of Ref.~\cite{Aaboud:2018xdl},   
where sub-jets with $R=0.2$ are removed from the large-$R$ jet 
if their transverse momentum is less than $5\%$ of the original large-$R$ jet transverse momentum. 
We generated events with $\sqrt{s} = 13$ TeV to recast the current experimental limit, 
and we use the same configuration to study future sensitivities at the HL-LHC, 
but the center of mass energy is set to $\sqrt{s} = 14$ TeV.

Figures~\ref{fig-efficiency13} and~\ref{fig-efficiency14} 
show the efficiencies of the sneutrino productions to the signal regions defined in Table~\ref{tab-monoW}
when $\sqrt{s} = 13$ TeV and $14$ TeV, respectively. 
$m_\tn > m_\te$ in the gray region. 
Since we assume that a half of events with the large-$R$ jet is classified into the HP regions, 
the efficiency factors of the LP regions are the same as that of the HP regions.

Events from the production with initial/final-state $W/Z$ boson tend to have smaller efficiencies, 
while those with the $W$ boson from the $\te$ decay tend to have larger efficiencies. 
The $p_T$ distribution of the latter is centered according 
to the mass difference between the charged slepton and sneutrino.  
In the merged topology signal regions, the efficiency increases as the mass difference increases.  
With fixed $m_\tn$, the efficiency decreases slightly as $m_\te$ increases, 
because the production of $\te\tn$ becomes less important 
against the production with initial/final state $W/Z$ boson. 
The $W/Z$ boson should be off-shell like to be categorized using the resolved topology; this is
so that the efficiency is maximized where the mass splitting is moderately large, on the order of $\sim 300$ GeV. 
No bottom quark is produced from $W$ boson decays.  
The efficiency to the signal regions with $b$-tagged jets are significantly small. 
Therefore the limit from the signal regions without $b$-tagged jets will give the most stringent limit.

\begin{table}[t!]
\centering
\caption{\label{tab-slims} 
The upper limits on the number of signal events in the signal regions. 
}
\begin{tabular}[t]{c|cccc|cc}\hline 
& 0b-HP    & 0b-LP    & 1b-HP    & 1b-LP     & 0b-Res & 1b-Res   \\ \hline\hline 
current                     & 205.1 & 218.3 & 89.25 & 122.2 & 887.6 & 318.9 \\
HL-LHC $2\sigma$ &1777   & 2181  & 782.1 & 929.7 & 8235 & 2869 \\
HL-LHC $5\sigma$ &4442  & 5453 & 1955 & 2324   & 20587 & 7174 \\ \hline 
\end{tabular}
\end{table}

The main SM backgrounds for these events originate 
from $t\bar{t}$ events and $Z+$jets events where the $Z$ decays invisibly to neutrinos. 
The search had found no excess on top of the backgrounds in 36.1 $\fbi$ of data. 
Thus we obtain the upper bounds on the signal events in each signal region 
from the experimental data and the expected number of SM backgrounds~\cite{Aaboud:2018xdl}.

We set the 95 $\%$ C.L. limit on the number of signal event $s$, such that 
\begin{align}
\mathrm{CL}_s := \frac{ \displaystyle \sum_{k=0}^{n}  \mathrm{Pois} \left( s+b| k \right) }
                                    { \displaystyle \sum_{k=0}^{n}  \mathrm{Pois} \left(b|k \right)  } = 0.05, 
\quad 
\mathrm{Pois}(\la|k) :=  \frac{\la^k e^{-\la}}{k!},   
\end{align}  
where $n$, $b$ are the number of observed events and backgrounds, respectively. 
We will also study the expected exclusion limit and discovery potential at the HL-LHC 
with the integrated luminosity of $3000$ $\fbi$. 
We simply assume that the number of backgrounds is scaled up by the integrated luminosity 
from the analysis with $36.1$ $\fbi$. 
The increase of the center-of-mass energy to 14 TeV, 
pile-up effects due to the high-luminosity environment, 
upgrades of the detector and so on, are neglected in this analysis. 
The signal events are generated with the same configuration as the $13$ TeV analysis, 
but the center of mass energy is set to $14$ TeV. 
The exclusion limit (discovery potential) at the HL-LHC on the number of signal events $s$ are estimated as 
\begin{align}
\frac{s}{\sqrt{b_{14\; \mathrm{TeV}}}}  = 2\; (5), 
\quad   
b_{14\; \mathrm{TeV}} = \frac{3000\; \fbi }{36.1\; \fbi} \times b_{13\; \mathrm{TeV}},  
\end{align}
where $b_{13\; \mathrm{TeV}}$ is the number of backgrounds at 13 TeV~\cite{Aaboud:2018xdl}.  
The upper limits of the signal events are summarized in Table.~\ref{tab-slims}.

\subsection{Results}
\label{sec-rslt}

\begin{figure}[t]
\centering
\includegraphics[width=0.8\textwidth]{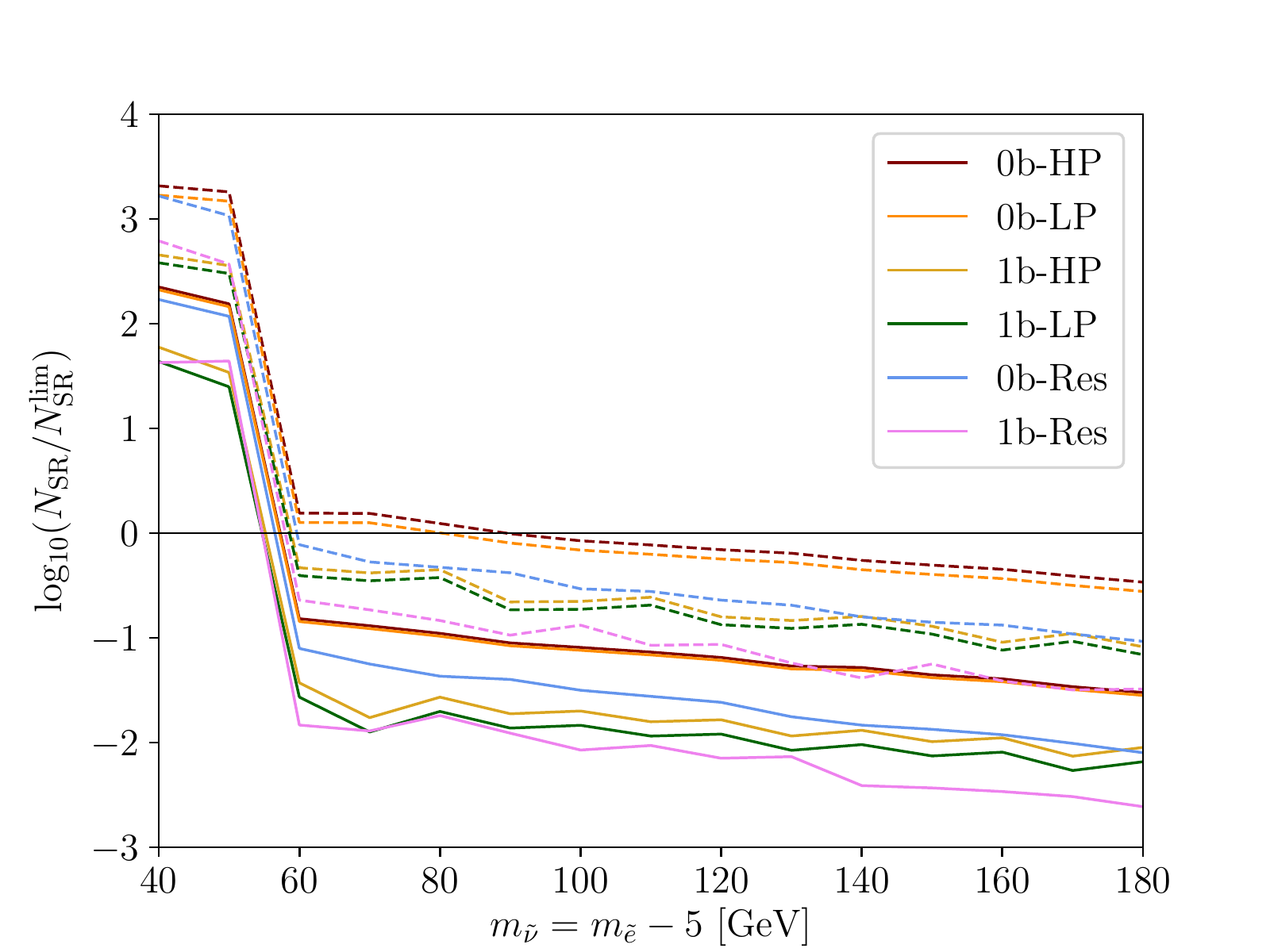}
\caption{
The experimental limits to the mass degenerate region.  
}
\label{fig:slepton_limit}
\end{figure}

Figure~\ref{fig:slepton_limit} shows the ratios of the number of events in the signal regions 
obtained by the simulation to the exclusion limits when the mass difference is 5 GeV.   
The solid lines are the current limit at 13 TeV, 
and the dashed lines are the $2\sigma$ limit at $14$ TeV with $3\;\mathrm{ab}^{-1}$ data. 
The colors represent the signal regions. 
The mass region is excluded if any of the lines are above the black line. 
All of the slepton pair production channels (namely $\tn\tn$, $\tn\te$, $\te\te$) are included in the analysis, 
since the decay products of the charged sleptons are too soft to be reconstructed.       
We see that the 0b-HP signal region is the most sensitive to slepton pair production, 
and the 0b-LP signal region has slightly weaker sensitivity. 
Although the efficiency of the 0b-Res topology is as large as that of the 0b-HP topology, 
the limit is weaker because of the large SM background.  
We see from Fig.~\ref{fig:slepton_limit} that the current limit on the sneutrino mass is about 55 GeV, 
and the expected limit from the full-run of the HL-LHC is about 90 GeV. 
The HL-LHC exclusion limit may be sensitive to the off-shell production of sneutrino pairs,   
but the discovery potential of $5\sigma$ is not sensitive to that and the limit is about 60 GeV.

\begin{figure}[t]
\centering
\includegraphics[width=0.8\textwidth]{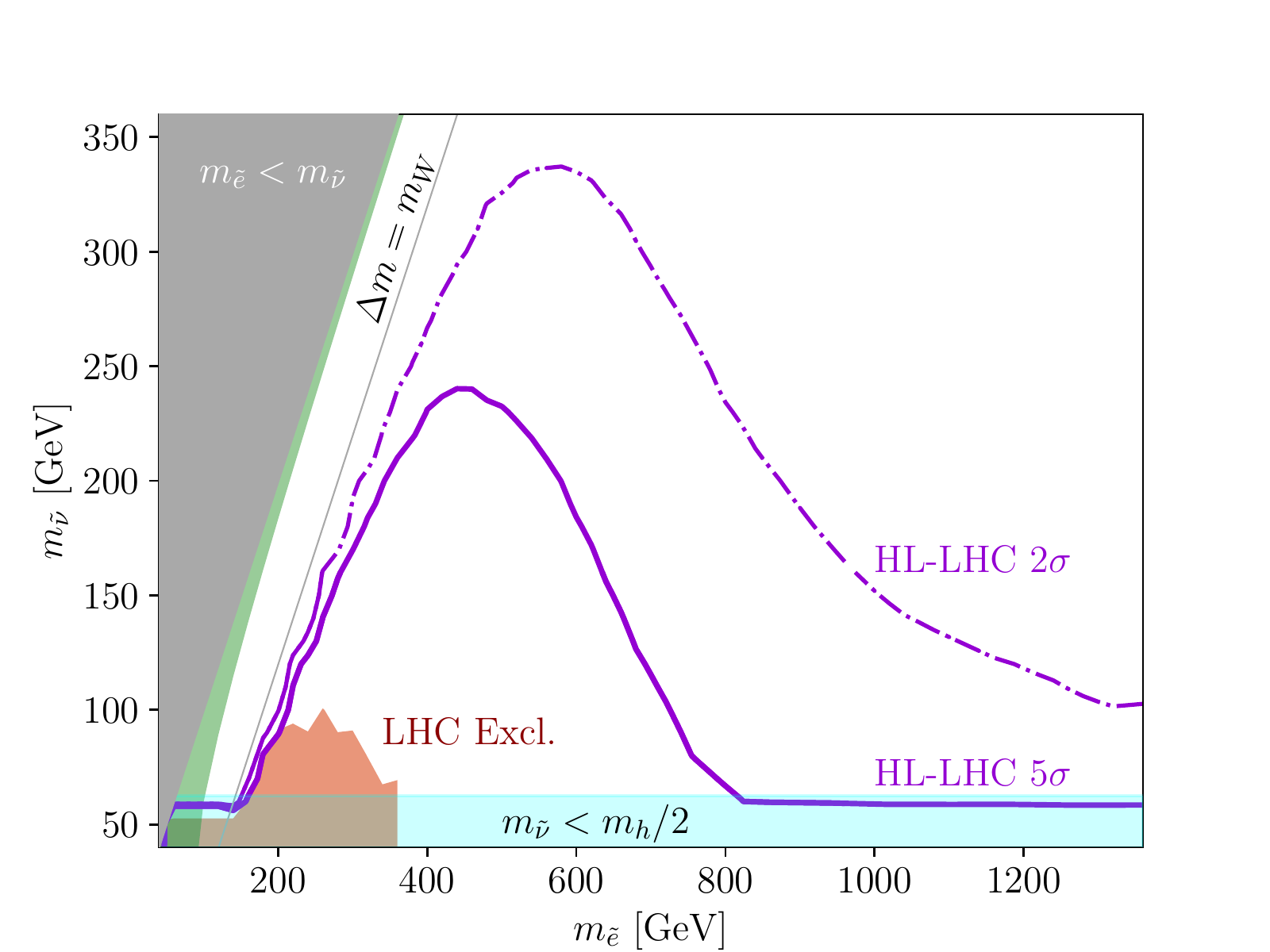}
\caption{Sneutrino-slepton parameter space, showing the constraints imposed by our analysis}
\label{fig:mass_plane-constraints}
\end{figure}

In Fig.~\ref{fig:mass_plane-constraints},  we see exclusions in the sneutrino--charged slepton mass plane.  
The gray region represents points where the charged slepton is lighter than the sneutrino 
and thus is not compatible with the sneutrino LSP scenario.  
The green region represents regions with mass splittings consistent with MSSM scenarios, 
with splittings given in Eq.~\eqref{eq-delm}.  
The sneutrinos are ligher than a half of the Higgs boson in the cyan region, 
where they may be excluded by the Higgs invisble decay unless the effective A-term $A_\tn$ 
is less than $\order{1}$ GeV. 
We have also placed a diagonal line to where $\Delta m = m_W$. 
In this figure, only the sneutrino pair production is included to obtain the limits at $\Delta m < m_W$. 
The productions involving a charged slepton may be irrelevant for the mono-$W/Z$ search
because the decay products of charged sleptons would change the topology of signals, 
unless these are too soft to be detected. 
The limits when charged sleptons are invisible at the detector have already shown in Fig.~\ref{fig:slepton_limit}. 
{
We expect that the limits would not change significantly for larger mass differences up to about 20 GeV. As the mass difference increases, the decays of charged sleptons will become visible. In particular, events with reconstructed leptons from off-shell $W$ bosons will be rejected by the requirement of the signal regions, see Table~\ref{tab-monoW}. This will reduce the sensitivity of the mono-$W/Z$ search, 
although the effect will not be drastic due to the small phase space and branching fraction of $W$ boson to leptons.  
}

The mono-$W$ becomes very powerful for a mass difference larger than $m_W$. 
In this region, the production process is dominated by charged slepton-sneutrino pair production, 
where the charged slepton decays through an on-shell $W$ in the final state. 
For these events, 
we may exclude sneutrinos in the 80--100 GeV range for charged sleptons between 200--300 GeV.  
However, once the charged slepton mass is sufficiently large, 
the production cross section drops, and the bound returns to the 55 GeV, 
which is the bound for production of sneutrino pairs alone.  
The purple curves show projections for the full 3~$\mathrm{ab}^{-1}$ run of the HL-LHC. 
The solid line shows the $5\sigma$ discovery potential in the mass plane. 
The dashed line gives a projection where
$s/\sqrt{b_{14\; \mathrm{TeV}}} = 2$ and constitutes a $\sim95\%$ C.L. exclusion assuming no excess is seen.  
In the MSSM region, 
the projected bound for HL-LHC is around 60 GeV 
if the decay products of the charged slepton is visible, 
while it is around 90 GeV if that is invisible as shown in Fig.~\ref{fig:slepton_limit}.

\section{Conclusions}
\label{sec-concl} 
The sneutrino (N)LSP scenario presents a challenge for discovering 
and constraining the slepton sector of supersymmetric models. 
We have used Higgs constraints and recast the ATLAS hadronic mono-boson search 
to bound slepton parameter space in light sneutrino scenarios.  
We see that 13 TeV data can place absolute lower bounds on sneutrinos of around 60 GeV 
an improvement over previous constraints from the LEP $Z$ invisible width. 
For more exotic regions of the sneutrino charged slepton mass plane 
we may place 80-100 GeV bounds on sneutrinos for charged slepton masses in the 200-300 GeV mass region.
Projections for the HL-LHC may place a lower mass bound of around 90 GeV for MSSM sneutrinos 
(N)LSP assuming no excess is seen.  
Conversely, if the light sneutrino scenario exists we may place the $5\sigma$ discovery potential 
is around $m_{\tilde{\nu}}\sim $60 GeV.

There are various further analyses that may be used 
in combination with our hadronic mono-$W/Z$ analysis to try to further improve the bounds 
and discovery potential of the sleptons in the light sneutrino scenario. 
For example a recasting of the monophoton analysis may also prove fruitful to zero in on this scenario. 
In addition, new analysis of events with soft leptons and missing energy have proven useful 
for constraining Higgsinos\cite{Aad:2019qnd}. 
Since we expect events which contain leptons of intermediate energy resulting 
from charged slepton decay into sneutrino in slepton pair production processes, 
we may imagine new analyses which take advantage of lepton(s) plus missing energy signals.

\section*{Acknowledgments}
The work of L.M.C, H.B.G. and J.K. 
is supported in part by the Department of Energy (DOE) under Award No.\ DE-SC0011726. 
This work of J.K. is supported in part by the Grant-in-Aid for Scientific Research from the
Ministry of Education, Science, Sports and Culture (MEXT), Japan No.\ 18K13534.

\end{document}